%% file: main.tex
\documentclass[twocolumn,times,tighten,astrosymb,trackchanges]{aastex631}
\input{affil}
\usepackage{CJK}
\usepackage{multirow}
\usepackage[caption=false]{subfig}
\newcommand\kms{~km~s$^{-1}$}
\newcommand\um{~$\mu$m}
\let\tablenum\relax
\usepackage{siunitx}
\usepackage{appendix}
\usepackage{amsmath,amsthm,amsfonts,amssymb,amscd}
\usepackage{booktabs}
\usepackage{tablefootnote}

\definecolor{maroon}{rgb}{0.760,0.118,0.337}

\definecolor{darkaqua}{rgb}{0.0,0.45,0.65}

\def\cm{\mbox{\,cm}}
\def\cm3{\mbox{\,cm$^{-3}$}}

\shorttitle{JWST SN~2021aefx}
\shortauthors{Kwok et al.}
\reportnum{published in ApJ Letters}

\begin{document}

\title{A \textit{JWST} Near- and Mid-Infrared Nebular Spectrum of the Type Ia Supernova 2021aefx}

\correspondingauthor{Lindsey A. Kwok}
\email{lindsey.kwok@physics.rutgers.edu}

\author[0000-0003-3108-1328]{Lindsey A.\ Kwok}
\Rutgers
\author[0000-0001-8738-6011]{Saurabh W.\ Jha}
\Rutgers
\author[0000-0001-7380-3144]{Tea Temim}
\Princeton
\author[0000-0003-2238-1572]{Ori D.\ Fox}
\STScI
\author[0000-0003-2037-4619]{Conor Larison}
\Rutgers
\author[0000-0002-9830-3880]{Yssavo Camacho-Neves}
\Rutgers
\author[0000-0002-8092-2077]{Max J.\ Brenner Newman}
\Rutgers
\author[0000-0002-2361-7201]{Justin D.\ R.\ Pierel}
\STScI
\author[0000-0002-2445-5275]{Ryan J.\ Foley}
\UCSC
\author[0000-0003-0123-0062]{Jennifer E.\ Andrews}
\GeminiNorth
\author[0000-0003-3494-343X]{Carles Badenes}
\UPitt
\ICCUB
\author[0000-0003-4769-4794]{Barnabas Barna}
\USzeged
\Prague
\author[0000-0002-4924-444X]{K.\ Azalee Bostroem}
\thanks{LSSTC Catalyst Fellow}
\UA
\author[0000-0001-8857-9843]{Maxime Deckers}
\UDublin
\author[0000-0003-2024-2819]{Andreas Fl\"ors}
\GSI
\author[0000-0003-4069-2817]{Peter Garnavich}
\NotreDame
\author[0000-0002-9154-3136]{Melissa L.\ Graham}
\DiRAC
\author[0000-0002-4391-6137]{Or Graur}
\ICG
\AMNH
\author[0000-0002-0832-2974]{Griffin Hosseinzadeh}
\UA
\author[0000-0003-4253-656X]{D.\ Andrew Howell}
\LCO
\UCSB
\author[0000-0002-8816-6800]{John P.\ Hughes}
\Rutgers
\author[0000-0001-5975-290X]{Joel Johansson}
\OKC
\author[0000-0002-7612-0469]{Sarah Kendrew}
\ESASTScI
\author[0000-0002-0479-7235]{Wolfgang E.\ Kerzendorf}
\MSUPA
\MSUCMSE
\author[0000-0003-2611-7269]{Keiichi Maeda}
\KyotoU
\author[0000-0002-9770-3508]{Kate Maguire}
\UDublin
\author[0000-0001-5807-7893]{Curtis McCully}
\LCO
\UCSB
\author[0000-0003-3615-9593]{John T.\ O'Brien}
\MSUPA
\author[0000-0002-4410-5387]{Armin Rest}
\STScI
\JHU
\author[0000-0003-4102-380X]{David J.\ Sand}
\UA
\author[0000-0002-9301-5302]{Melissa Shahbandeh}
\STScI
\author[0000-0002-7756-4440]{Louis-Gregory Strolger}
\STScI
\author[0000-0003-4610-1117]{Tam\'as Szalai}
\USzeged
\ELKHSZTE
\Konkoly
\author[0000-0002-5221-7557]{Chris Ashall}
\VTech
\author[0000-0001-5393-1608]{E. Baron}
\UOklahoma
\Hamburg
\author[0000-0003-4625-6629]{Chris R.\ Burns}
\Carnegie
\author[0000-0002-7566-6080]{James M.\ DerKacy}
\VTech
\author[0000-0001-5888-2542]{Tyco Mera Evans}
\FSU
\author[0000-0002-5253-3584]{Alec Fisher}
\FSU
\author[0000-0002-1296-6887]{Llu\'is Galbany}
\ICE
\IEEC
\author[0000-0002-4338-6586]{Peter Hoeflich}
\FSU
\author[0000-0003-1039-2928]{Eric Hsiao}
\FSU
\author[0000-0001-6069-1139]{Thomas de Jaeger}
\IfA
\author[0000-0001-6209-838X]{Emir Karamehmetoglu}
\Aarhus
\author[0000-0002-6650-694X]{Kevin Krisciunas}
\TAMU
\Mitchell
\author[0000-0001-8367-7591]{Sahana Kumar}
\FSU
\author[0000-0002-3900-1452]{Jing Lu}
\FSU
\author[0000-0003-0733-7215]{Justyn Maund}
\USheffield
\author[0000-0001-6876-8284]{Paolo A.\ Mazzali}
\Liverpool
\MaxPlanck
\author[0000-0001-7186-105X]{Kyle Medler}
\Liverpool
\author[0000-0003-2535-3091]{Nidia Morrell}
\LasCampanas
\author[0000-0003-2734-0796]{Mark. M.\ Phillips}
\LasCampanas
\author[0000-0003-4631-1149]{Benjamin J.\ Shappee} 
\IfA
\author[0000-0002-5571-1833]{Maximilian Stritzinger}
\Aarhus
\author[0000-0002-8102-181X]{Nicholas Suntzeff}
\TAMU
\Mitchell
\author[0000-0002-0036-9292]{Charles Telesco}
\UFlorida
\author[0000-0002-2471-8442]{Michael Tucker}
\thanks{CCAPP Fellow}
\OSU
\author[0000-0001-7092-9374]{Lifan Wang}
\TAMU
\Mitchell

\begin{abstract}

We present \textit{JWST} near- and mid-infrared spectroscopic observations of the nearby normal Type Ia supernova SN~2021aefx in the nebular phase at $+255$ days past maximum light. Our Near Infrared Spectrograph (NIRSpec) and Mid Infrared Instrument (MIRI) observations, combined with ground-based optical data from the South African Large Telescope (SALT), constitute the first complete optical$+$NIR$+$MIR nebular SN Ia spectrum covering 0.3--14\um. This spectrum unveils the previously unobserved 2.5$-$5\um\ region, revealing strong nebular iron and stable nickel emission, indicative of high-density burning that can constrain the progenitor mass. The data show a significant improvement in sensitivity and resolution compared to previous \textit{Spitzer} MIR data. We identify numerous NIR and MIR nebular emission lines from iron-group elements as well as lines from the intermediate-mass element argon. The argon lines extend to higher velocities than the iron-group elements, suggesting stratified ejecta that are a hallmark of delayed-detonation or double-detonation SN Ia models. We present fits to simple geometric line profiles to features beyond 1.2\um\ and find that most lines are consistent with Gaussian or spherical emission distributions, while the [Ar III] 8.99\um\ line has a distinctively flat-topped profile indicating a thick spherical shell of emission. Using our line profile fits, we investigate the emissivity structure of SN~2021aefx and measure kinematic properties. Continued observations of SN~2021aefx and other SNe Ia with \textit{JWST} will be transformative to the study of SN Ia composition, ionization structure, density, and temperature, and will provide important constraints on SN Ia progenitor and explosion models.

\end{abstract}


\section{Introduction\label{sec:intro}}

Type Ia supernovae (SN Ia) play an important role in astrophysics and cosmology, yet we still lack a detailed understanding of their progenitor systems and explosion physics. Nebular phase spectroscopy at late times \citep[beyond about 100 days past maximum light;][]{Bowers1997, Branch2008, Silverman2013, Friesen2014, Black2016} reveals the SN ejecta when they have expanded, allowing us to see to the innermost material. The observed flux is dominated by optically-thin forbidden-line emission that directly probes the composition, density, temperature, and ionization structure of the ejecta, constraining models of thermonuclear explosions of a white dwarf \citep[for a review, see][]{jerkstrand_spectra_2017}.

At early times, most SN Ia flux is at optical wavelengths, but as the ejecta fade and cool to the nebular phase, the near- and mid-infrared (NIR and MIR) comprise a large fraction of the emission \citep{Axelrod1980, Fransson2015}. Nebular spectra have been obtained for hundreds of SNe Ia in the optical but far fewer in the ground-accessible NIR windows. There are only three SNe Ia to date with published nebular spectra in the MIR: one epoch each of SN 2003hv and SN 2005df, observed with \textit{Spitzer} ($\lambda =$ 5--15\um\ at spectral resolution R $\sim$ 90; \citealt{Gerardy2007}) and four epochs of SN 2014J, observed from the ground with Gran Telescopio Canarias (GTC) ($\lambda =$ 8--13\um\ with R $\sim$ 60; \citealt{Telesco2015}) at phases of $+$57, $+$81, $+$108, and $+$137 days. Atmospheric absorption and sky background limit ground-based capabilities at these wavelengths. \textit{Spitzer} was pushed to its sensitivity limits and useful observations of these three SNe Ia were only possible because they were nearby ($d \sim$ 3.5 Mpc for SN 2014J \citep{Dalcanton2009, Goobar2014}, and $d \lesssim $ 20 Mpc for SN 2003hv and SN 2005df \citep{Gerardy2007}).

Nebular phase observations in the NIR and MIR provide unique and powerful constraints on models, including the density-dependent nucleosynthesis of intermediate mass elements, radioactive iron-group elements, and stable iron-group elements \citep{Gerardy2007, Diamond2018, Dhawan2018, Hoeflich2021}. The JWST Near Infrared Spectrograph \citep[NIRSpec;][]{Jakobsen2022} and Mid Infrared Instrument \citep[MIRI;][]{Rieke2015, Wright2015} provide access to a wider range of elemental and ionic species than the optical. Lines are also typically less blended in the infrared, making it easier to derive line fluxes and abundance estimates, as well as to infer the geometry of the emission region from the line profile shape \citep[e.g.,][]{jerkstrand_spectra_2017}. Infrared spectra can also show evidence of dust formation in the SN ejecta and polycyclic aromatic hydrocarbon (PAH) line features that reveal the local and galactic environment \citep{Tielens2005, Wang2005, Rho2008, Johansson2017}.

\textit{JWST}, with its wider NIR and MIR wavelength coverage, better spectral resolution, and enormously increased sensitivity compared to previous facilities, will be transformative to our understanding of SNe Ia. Here we present the first NIR $+$ MIR SN Ia spectrum from \textit{JWST}, of SN~2021aefx, covering 0.6--14\um. Our data include the previously unobserved 2.5--5\um\ region and, combined with ground-based optical data, create a continuous optical $+$ NIR $+$ MIR SN Ia spectrum. This observation, taken as part of \textit{JWST} Cycle 1 GO program 2072 ``See Through Supernovae'' (PI: S. W. Jha), is the initial epoch of the first SN in a program to build a legacy, reference sample of \textit{JWST} NIR $+$ MIR nebular spectra of 9 white-dwarf (thermonuclear) SNe over three cycles. SN~2021aefx will also be the target of two epochs of data from the \textit{JWST} Cycle 1 GO program 2114 (PI: C. Ashall). Combined, these programs will observe SN~2021aefx in four epochs, providing the most comprehensive time series of nebular IR spectra for a SN Ia.

SN~2021aefx is a ``normal" \citep[see e.g. ][]{Blondin2012} SN Ia that was discovered within hours of explosion by the Distance Less Than 40~Mpc (DLT40) survey \citep{tartaglia_early_2018} on 2021 November 11.3 UT at $\alpha=04\textsuperscript{h}19\textsuperscript{m}53\fs400$ and $\delta=-54\degr56'53\farcs09$ (J2000) \citep{Hosseinzadeh2022}. It is located in the nearby galaxy NGC~1566 with a distance of 18.0 $\pm$ 2.0 Mpc \citep[$\mu =$ 31.28 $\pm$ 0.23 mag;][]{sabbi_resolved_2018} and a redshift of $z=0.005017$ \citep{allison_search_2014}, making it a bright target for our first observation with \textit{JWST}. 

SN~2021aefx peaked at an apparent magnitude of \textit{B} $=$ 11.7~mag \citep[$M_B = -19.6$~mag;][]{Hosseinzadeh2022} and exhibited an exceptionally high silicon velocity ($v \simeq$ 30,000\kms) in the earliest spectrum \citep{Bostroem2021}. High-cadence intra-night observations of SN~2021aefx were carried out by the Precision Observations of Infant Supernova Explosions \citep[POISE;][]{Burns2021, ashall_speed_2022} and the DLT40 surveys, and additional multi-wavelength follow-up photometry revealed an early light-curve excess, perhaps attributable to interaction between the ejecta and a nondegenerate companion star, interaction with circumstellar material, or the effect of an unusual nickel distribution \citep{Hosseinzadeh2022}. Part of the early light curve evolution may also be explained by high and rapidly-evolving ejecta velocities \citep{ashall_speed_2022}. Aside from this peculiarity at the earliest epochs (which have rarely been probed), SN~2021aefx subsequently evolved into a normal SN Ia that would be included in cosmological samples. The evolution of SN~2021aefx has been closely followed by ground-based observatories and has generated significant interest in the SN community.

In \autoref{sec:obs} we detail our observations and data reduction; in \autoref{sec:lines} we identify optical, NIR, and MIR nebular emission lines in SN~2021aefx; and in \autoref{sec:profiles} we present basic geometric line profile fits to the dominant spectral features. We discuss the implications of our results and conclude in \autoref{sec:conclusions}.

\begin{figure*}
    \centering
    \includegraphics[width=\textwidth]{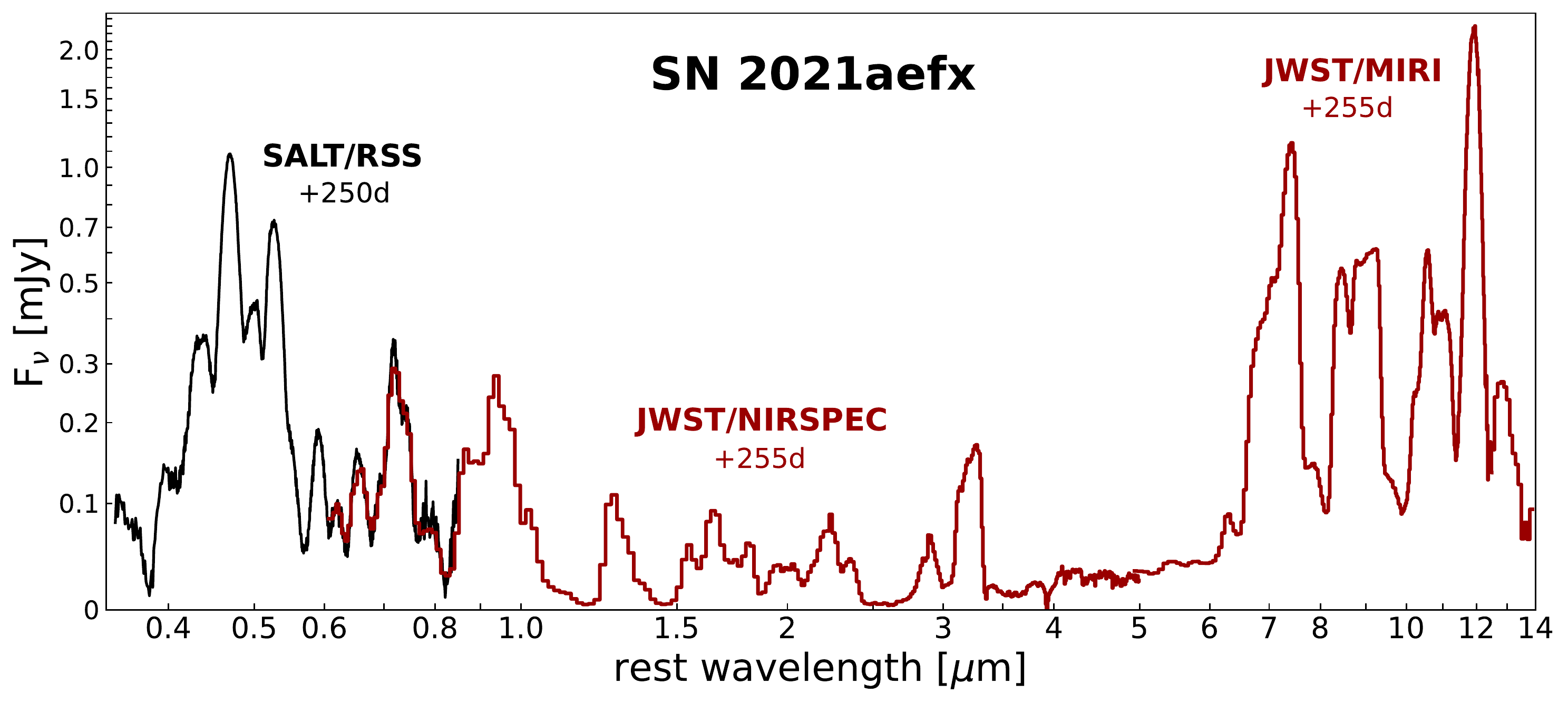}
    \caption{Combined optical $+$ NIR $+$ MIR spectrum of Type Ia SN~2021aefx in the nebular phase. The optical data are from SALT/RSS and the NIR and MIR data were obtained with \textit{JWST}/NIRSpec and \textit{JWST}/MIRI, respectively. The optical flux is calibrated to ground-based photometry and the MIRI flux is scaled to the MIRI F1000W photometry. The NIRSpec flux is unscaled from the \textit{JWST} pipeline and matches up well to the optical and MIR. The spectrum has been dereddened and corrected for the host-galaxy redshift. For presentation purposes, the optical spectrum and the MIR spectrum past 12.5\um\ have been rebinned to lower resolution. The flux axis uses a non-linear (arcsinh) scale to better show all the features across a wide range of wavelength and $F_\nu$. All subsequent spectra presented in the paper use a linear flux scale.}
    \label{fig:full_spec}
\end{figure*}

\section{Observations \label{sec:obs}}

We present the \textit{JWST} spectrum of SN~2021aefx in \autoref{fig:full_spec}. We observed SN~2021aefx with both NIRSpec in the Fixed Slits (FS) Spectroscopy mode \citep{Jakobsen2022, Birkmann2022, Rigby2022} with the prism and MIRI in the Low Resolution Spectroscopy (LRS) mode \citep{Kendrew2015, Kendrew2016, Rigby2022} on 2022 August 11.7 UT at a rest-frame phase of $+$254.9d, relative to \textit{B}-band maximum \citep[59546.54 $\pm$ 0.03 MJD;][]{Hosseinzadeh2022}.

Our NIRSpec observations used the S200A1 (0.2\arcsec\ wide $\times$ 3.3\arcsec\ long) slit with the PRISM grating and CLEAR filter and our MIRI observations used the LRS slit with the P750L disperser. The combined wavelength coverage spans 0.6--14\um. Details of the observation settings are given in \autoref{tab:details}.

\begin{figure}
    \centering
    \includegraphics[width=0.45\textwidth]{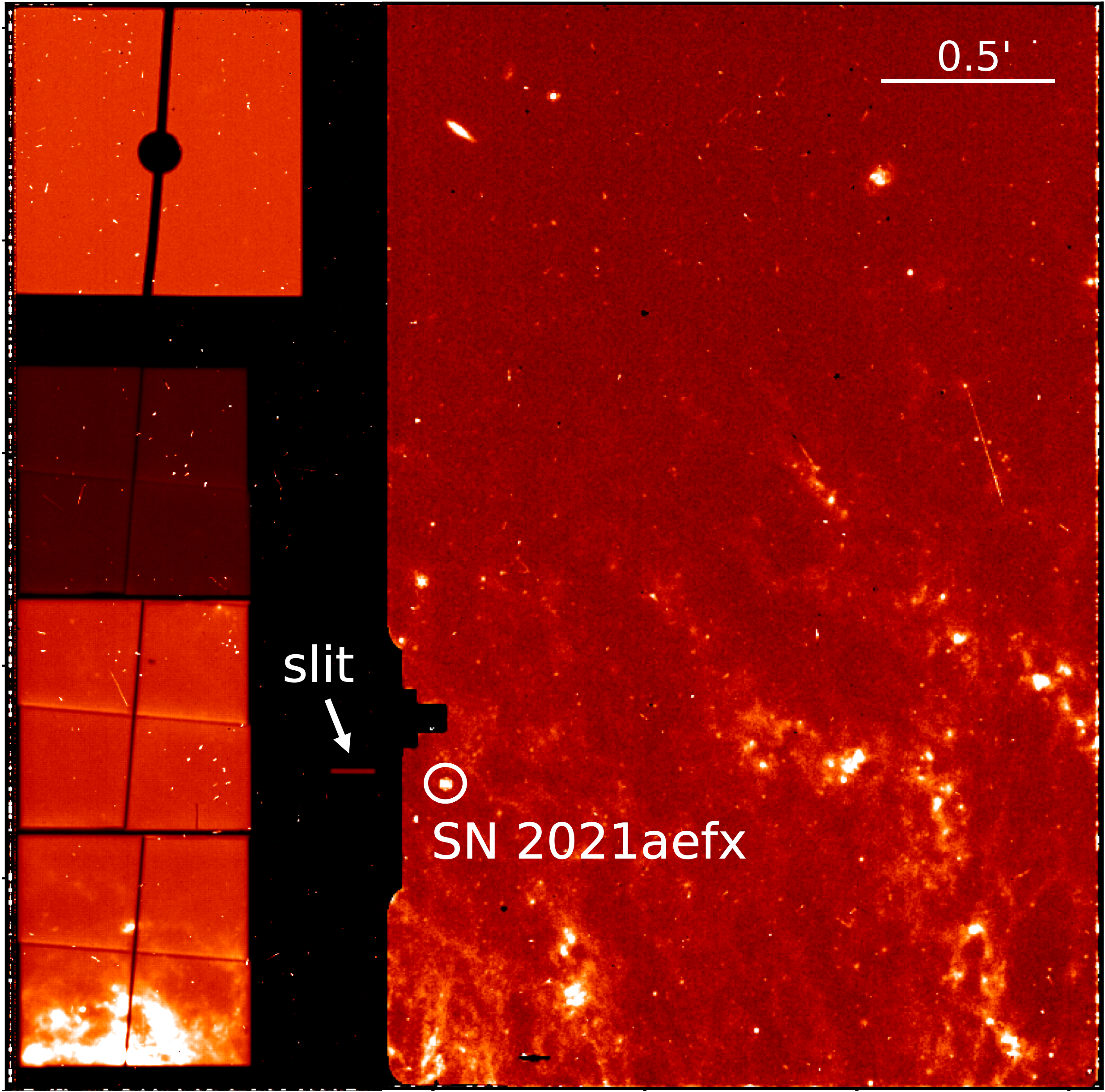}
    \caption{\textit{JWST} MIRI F1000W verification image showing SN~2021aefx during target acquisition, before placement on the LRS slit. Part of the central region of the host galaxy, NGC~1566, can be seen in the bottom left corner; gas and dust features can be seen in the image on the right. Cosmic rays have been removed from the image.}
    \label{fig:miri_image}
\end{figure}

\begin{deluxetable}{lll}[t]
\tablecaption{JWST SN~2021aefx Observation Details\label{tab:details}}
\tablehead{\colhead{Setting} & \colhead{NIR} & \colhead{MIR}}
\tablecolumns{3}
\startdata
Instrument & NIRSpec & MIRI \\
Mode & FS & LRS \\
Wavelength Range & 0.6 $-$ 5.3\um\ & 5 $-$ 14\um\ \\
Slit & S200A1 (0.2\arcsec\ x 3.3\arcsec) & Slit \\
Grating/Filter & PRISM/CLEAR & P750L \\
R $=\lambda/\Delta\lambda$ & $\sim$100 & $\sim$40 -- 250 \\
Subarray & FULL & FULL \\
Readout Pattern & NRSIRS2RAPID & FASTR1 \\
Groups per Integration & 5 & 134 \\
Integrations per Exposure & 2 & 2 \\
Exposures/Nods & 3 & 2 \\
Total Exposure Time & 525 s & 1493 s \\
Target Acq. Exp. Time & 4 s & 89 s
\enddata
\end{deluxetable}

\subsection{\textit{JWST} Data Reduction}

The data were reduced using the publicly available ``jwst''\footnote{\url{https://github.com/spacetelescope/jwst}} pipeline \citep[version 1.8.0;][]{Bushouse_JWST_Calibration_Pipeline_2022} routines for bias and dark subtraction, background subtraction, flat field correction, wavelength calibration, flux calibration, rectification, outlier detection, resampling, and spectral extraction. The final NIRSpec ``stage 3'' one-dimensional (1D) spectrum extracted by the automated pipeline, available on the Mikulski Archive for Space Telescopes (MAST)\footnote{\url{https://mast.stsci.edu/portal/Mashup/Clients/Mast/Portal.html}}, was of sufficiently good quality that we did not rerun any portion of the pipeline. Unfortunately, the MIRI stage 3 1D spectrum extracted by the automated pipeline, available on MAST, was noisy and unsuitable because the automated extraction aperture was not properly centered on the SN trace. Thus, we re-extracted the spectrum by manually running stage 3 of the pipeline (calwebb\_spec3) from the stage 2 (calwebb\_spec2) data products, enforcing the correct extraction trace and aperture.

We identified an issue with the original MIRI/LRS slit wavelength calibration from the \textit{JWST} pipeline that has been noted and confirmed by others and is under investigation (S.~Kendrew \& G.~Sloan, private communication). Spectra of the candidate Herbig B[e] star VFTS~822, which exhibits hydrogen emission lines \citep{Kalari2014}, were taken as part of  calibration programs \textit{JWST} Cycle 0 COM/MIRI 1259\footnote{\url{https://www.stsci.edu/jwst/phase2-public/1259.pdf}} (PI: S. Kendrew) and \textit{JWST} Cycle 1 CAL/MIRI 1530\footnote{\url{https://www.stsci.edu/jwst/phase2-public/1530.pdf}} (PI: S. Kendrew). From the data publicly available on the MAST archive, we measured wavelength centroids and uncertainties for 12 H I emission line peaks in the data and matched them with their known rest wavelengths. We found significant offsets between the wavelengths from the pipeline calibration and the H I emission line wavelengths, with larger deviation at shorter wavelengths ($\sim$0.1\um\ at the longest wavelengths, rising to $\sim$0.5\um\ at the shortest wavelengths). We developed a custom wavelength solution correction that informed updates to the \textit{JWST} pipeline by the MIRI Team. All MIRI/LRS slit data on MAST have been reprocessed by the new wavelength calibration, including the MIRI data of SN~2021aefx presented in this work. The inaccuracy in the wavelength calibration has been reduced to $\sim$0.02 -- 0.05\um\ (MIRI Team, private communication). We caution that the H I emission line peaks in the VFTS~822 data are weak and difficult to fit; full resolution of this wavelength calibration issue may require additional observations of other sources.

We measured MIRI F1000W photometry of SN~2021aefx from the LRS verification image (see \autoref{fig:miri_image}) with $F_\nu =$ 0.309 $\pm$ 0.010 mJy, corresponding to 17.67 $\pm$ 0.04 mag AB. The photometry was done on the F1000W data from the \textit{JWST} pipeline using a 70\% encircled energy aperture radius (4.3 pixels) and inner and outer sky radii of 6.063 and 10.19 pixels (and a corresponding aperture correction was also applied). Integrating the MIRI spectrum of the supernova over the F1000W passband gave a flux that agreed with the measured photometry to within 2\%. We applied a rescaling of the spectrum to match the photometry precisely. The NIRSpec spectrum does not have similarly measured photometry, but the pipeline spectrum matches both the optical and MIR spectra well, so we do not adjust its flux calibration.

To correct for extinction by dust, we use the Python \texttt{dust-extinction} package \citep[v.~1.1;][]{karl_gordon_2022_6397654}. We deredden the NIRSpec spectrum out to 1.0\um\ using the F19 model from \citet{Fitzpatrick2019} and the NIRSpec spectrum past 1.0\um\ as well as the MIRI spectrum with the G21\_MWavg model from \citet{Gordon2021}. We correct for both the host-galaxy extinction of \textit{E(B$-$V)}$_\text{host} =$ 0.097 mag \citep{Hosseinzadeh2022} and the Milky Way extinction of \textit{E(B$-$V)}$_\text{MW} =$ 0.008 mag \citep{schlafly_measuring_2011}.

\subsection{Optical Data}

We obtained an optical nebular spectrum of SN~2021aefx with the Southern African Large Telescope (SALT) Robert Stobie Spectrograph \citep[RSS; ][]{RSS} on 2022 August 7.1 UT (rest-frame phase of $+$250.3 d) using a 1\farcs5 longslit and the PG0900 grating in four tilt settings with a total exposure time of 2294 s. Using a custom pipeline based on standard Pyraf \citep{Pyraf} spectral reduction routines and the PySALT package \citep{PySALT}, we reduced the optical spectrum and removed cosmic rays, host galaxy lines and continuum, and telluric absorption. We scaled the optical spectrum to observed contemporaneous \textit{UBgVri} photometry, obtained with the Sinistro cameras on Las Cumbres Observatory's 1\,m telescopes \citep{brown_cumbres_2013} and reduced automatically by the BANZAI \citep{mccully_lcogt/banzai:_2018} and \texttt{lcogtsnpipe} \citep{valenti_diversity_2016} pipelines, using the \texttt{speccal} module in the Light Curve Fitting package \citep{hosseinzadeh_light_2020}. Lastly, we applied a redshift correction to the host-galaxy rest frame and dereddened using the F19 model \citep{Fitzpatrick2019} as above.

\begin{figure}
    \centering
    \includegraphics[width=0.5\textwidth]{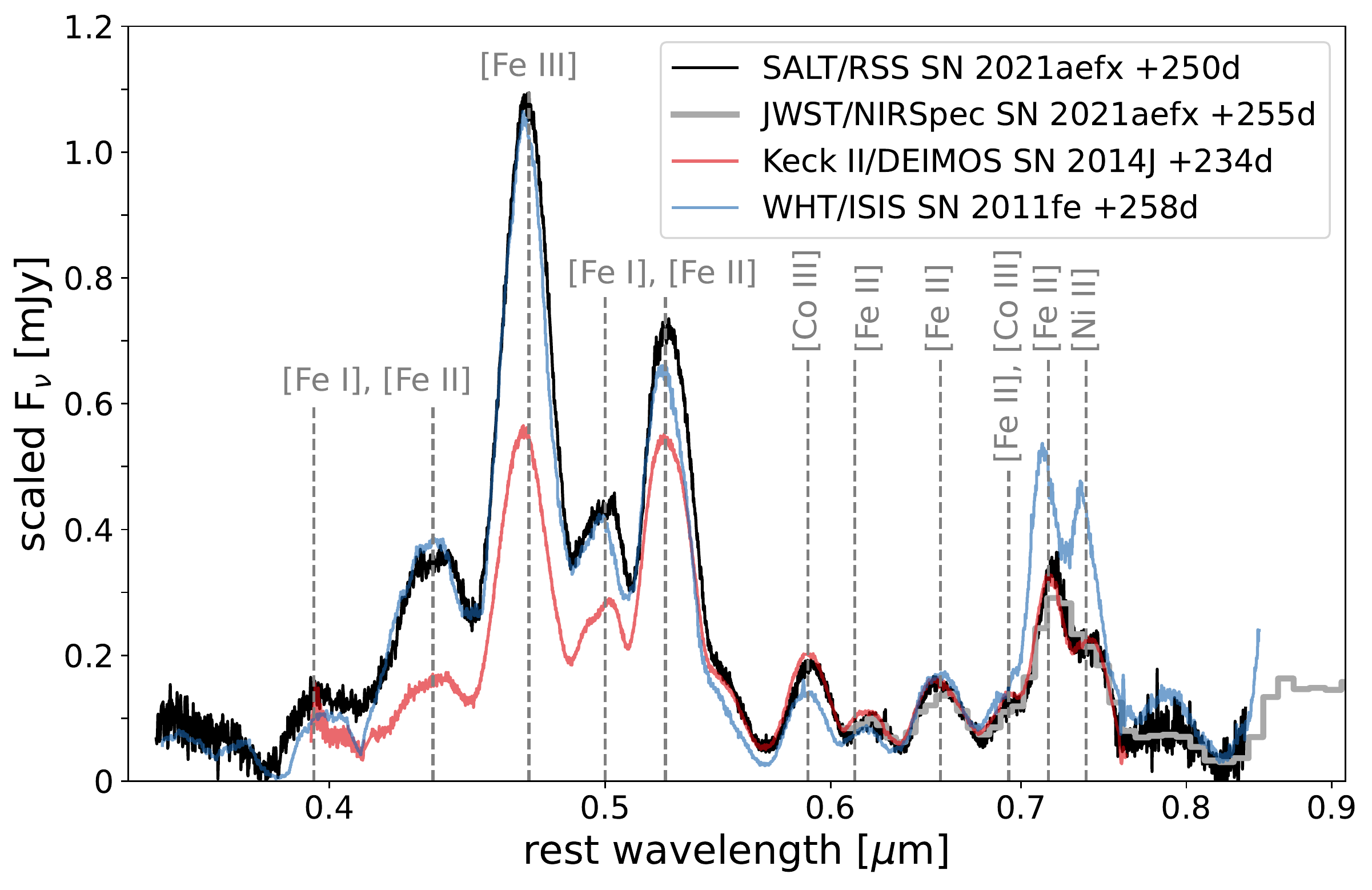}
    \caption{Optical line identifications for SN~2021aefx; only the most dominant species for each feature are shown. The optical spectrum (black) from SALT/RSS on 2022 August 7 at a rest-frame phase of $+$250d is dominated by forbidden-line emission from iron-group elements. For comparison, we plot the \textit{JWST}/NIRSpec spectrum (grey) on 2022 August 11 at $+$255d, and the optical spectra of SN~2011fe \citep[blue;][]{Mazzali2015} and SN~2014J \citep[red;][]{Childress2015} at similar phase ($+$258d and $+$234d, respectively). Phases are given with respect to \textit{B}-band maximum and all spectra have been dereddened.}
    \label{fig:optical_ID}
\end{figure}

\section{Line Identification\label{sec:lines}}

We identify nebular emission lines in the optical, NIR, and MIR spectra presented in this work using line identifications from the Atomic Line List\footnote{\url{https://www.pa.uky.edu/~peter/newpage/}}; the Atomic-ISO line list\footnote{\url{https://www.mpe.mpg.de/ir/ISO/linelists/FSlines.html}}; previous optical line identifications by \citet{Graham2022}, \citet{Tucker2022}, and \citet{Wilk2020}; previous NIR line identifications by \citet{Diamond2018}, \citet{Dhawan2018}, \citet{Hoeflich2021}, and \citet{Mazzali2015}; previous MIR line identifications by \citet{Gerardy2007} and \citet{Telesco2015}; and optical $+$ NIR $+$ MIR lines from models by \citet{Flors2020}.

In this work, we focus on line identifications for the 2.5--5\um\ region in the NIR and the full 5$-$14\um\ MIR. Candidate lines in these wavelength regions of interest were narrowed down by selecting atomic species consistent with SN Ia abundance models \citep[e.g. ][]{Nomoto1984, Thielemann1986, Fink2010, Pakmor2012, Seitenzahl2013}, predicted strength of the forbidden-line transitions, and proximity to ground-state transitions. 

Selected prominent optical lines are marked in \autoref{fig:optical_ID}. The optical lines match closely with those presented by \citet{Graham2022}, \citet{Tucker2022}, and \citet{Wilk2020}, and are dominated by blended emission lines from the iron-group elements: [Fe II], [Fe III], [Co III], and [Ni II]. We compare to an optical spectrum of SN~2011fe from \citet{Mazzali2015} using the William Herschel Telescope (WHT) with the Intermediate-dispersion Spectrograph and Imaging System (ISIS) and dereddened by \textit{E(B$-$V)} $=$ 0.023 mag. We also compare to a spectrum of SN~2014J from \citet{Childress2015} taken with the Keck II telescope and the DEep Imaging Multi-Object Spectrograph (DEIMOS), dereddened by $R_V$ = 1.4, \textit{E(B$-$V)} $=$ 1.2 mag \citep{Foley2014, Amanullah2014, Mazzali2015}. The optical spectrum and line identification of SN~2021aefx closely matches that of SN~2011fe and SN~2014J, indicating that SN~2021aefx is representative of a typical SN Ia at about $+$250d. The spectrum of SN~2011fe is slightly blueshifted compared to SN~2021aefx.

\subsection{NIR Emission Lines \label{sec:nir_lines}}

\begin{figure*}
    \centering
    \includegraphics[width=\textwidth]{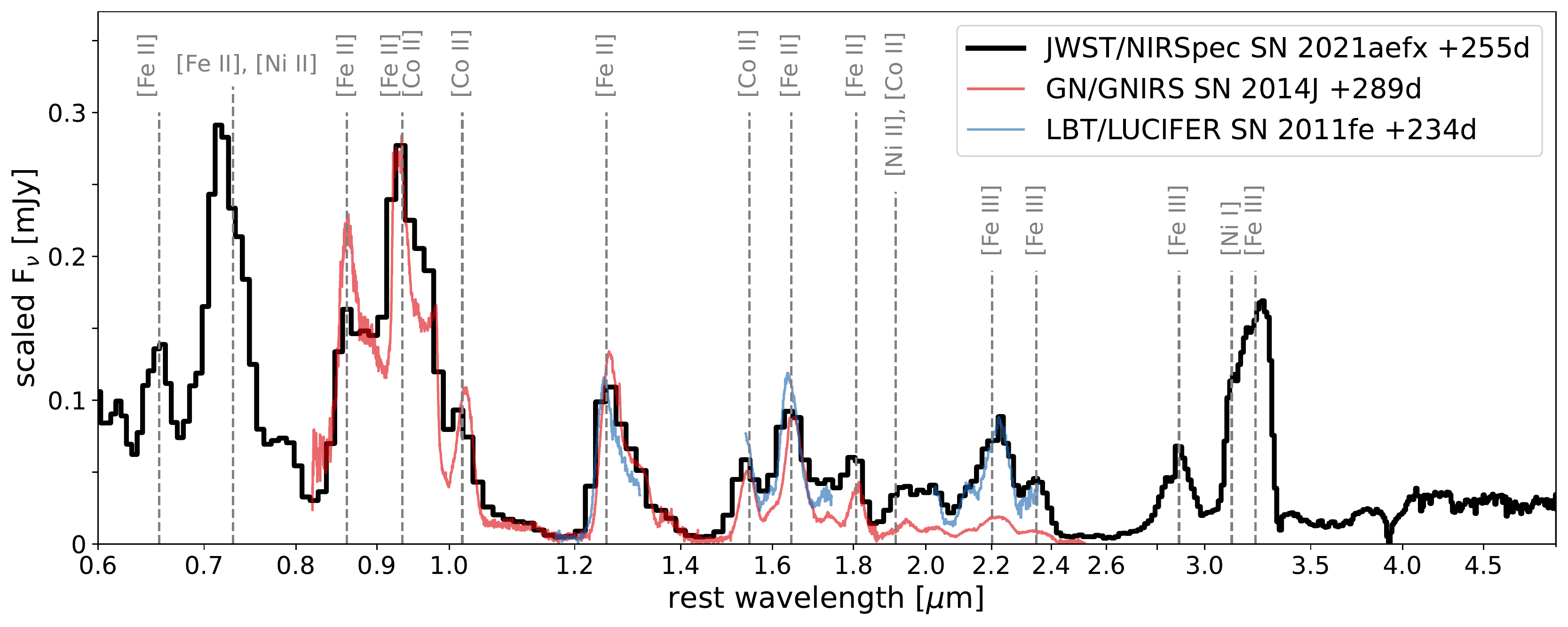}
    \caption{NIR line identifications for SN 2021aefx; only the most dominant species for each feature are shown. We compare the \textit{JWST}/NIRSpec spectrum of SN~2021aefx (black) at $+$255d with a scaled GN/GNIRS spectrum of SN~2014J (red) at the closest available phase of $+$289d \citep{Diamond2018, Graur2020} and a scaled LBT/LUCIFER \textit{JHK}$_\text{s}$ spectrum of SN~2011fe (blue) at a phase of $+$234d \citep{Mazzali2015}. Nearly all emission lines are from significantly blended iron-group elements. The 2.5$-$5\um\ region shows two strong blended features of predominantly [Fe III] 2.874 \& 2.905\um, and [Ni I] 3.120\um\ \& [Fe III] 3.229\um, and several weak features. All spectra have been dereddened. Details of the NIR lines can be found in \autoref{tab:lines}.}
    \label{fig:nirspec_ID}
\end{figure*}

In \autoref{fig:nirspec_ID}, we mark prominent lines in the \textit{JWST}/NIRSpec spectrum of SN~2021aefx. Most of the lines in the NIR are considerably blended, so for clarity we only mark the most dominant species for each feature \citep{Diamond2018, Dhawan2018, Mazzali2015, Hoeflich2021}. A more comprehensive set of identifications and a detailed list of strong NIR transitions from 0.8--2.4\um\ can be found in Table 2 and associated figures from \citet{Diamond2018} and additional model line transitions in this region are given by \citet{Flors2020} and \citet{Hoeflich2021}. 

Comparison to a dereddened \citep[$R_V$ = 1.4, \textit{E(B$-$V)} $=$ 1.2 mag;][]{Foley2014, Amanullah2014, Mazzali2015}, scaled Gemini North GNIRS \citep{Elias2006a, Elias2006b} spectrum of SN~2014J at the closest available phase of $+$289d \citep{Diamond2018, Graur2020} and a dereddened \citep[\textit{E(B$-$V)}$=$0.023 mag;][]{Mazzali2015}, scaled Large Binocular Telescope LUCIFER \citep{Seifert2003} spectrum of SN~2011fe at similar phase of $+$234d shows good agreement between the SNe, with nearly all of the same lines present. The line strengths vary, though this may be attributed to the differences in phase. The NIR spectral features are predominantly blends of forbidden-line emission from the iron-group elements Fe, Co, and Ni. The [Ni I], [Ni II], and [Ni III] transitions are of particular interest for constraining models; however, none of the NIR nickel lines are isolated.

\begin{figure}
    \centering
    \includegraphics[width=0.5\textwidth]{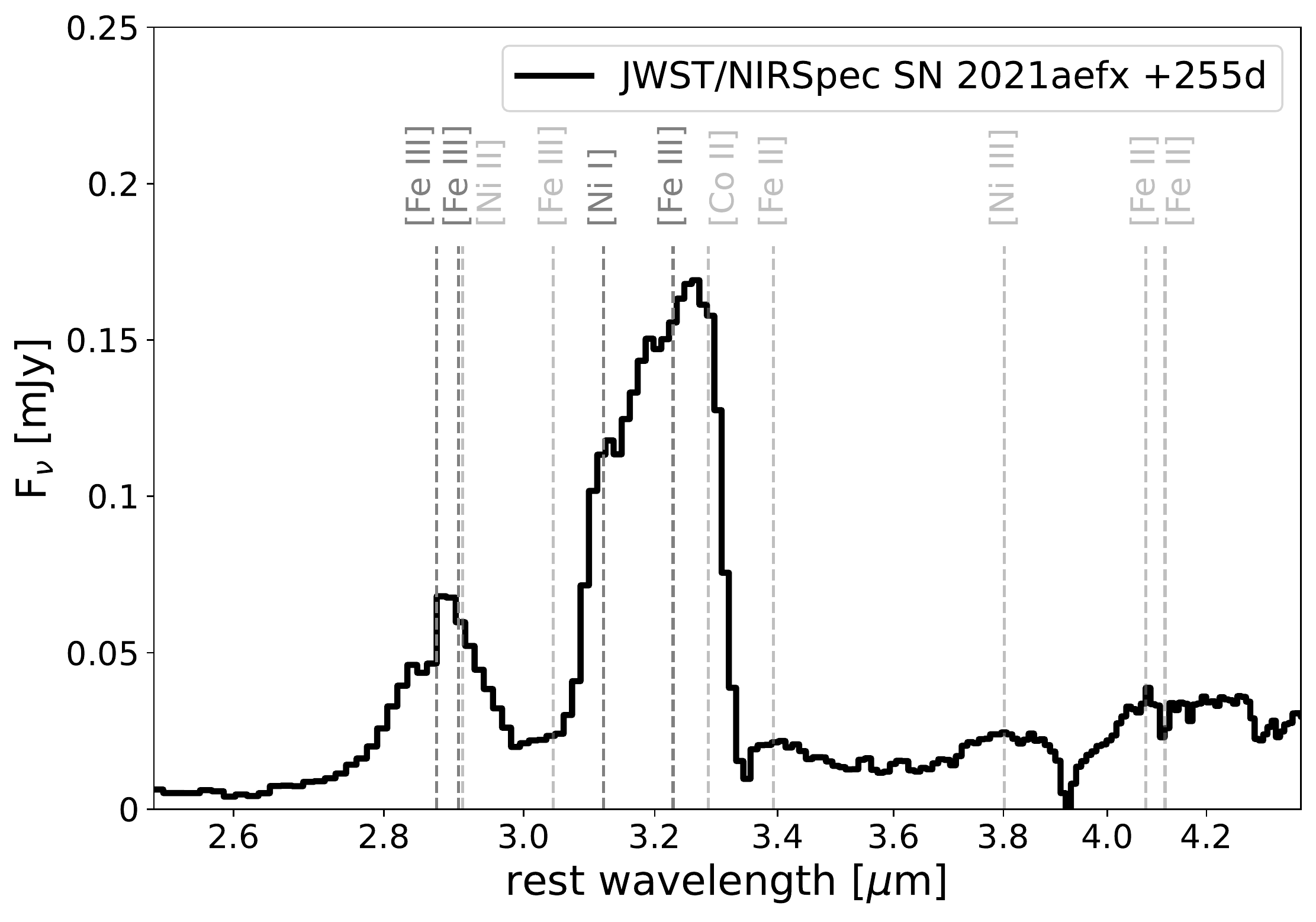}
    \caption{\textit{JWST}/NIRSpec line identifications in the 3\um\ region, previously unobserved for SN Ia. The main line features arise from [Fe III] 2.874 \& 2.905\um, and [Ni I] 3.120\um\ \& [Fe III] 3.229\um\ emission (grey). Several possible weaker lines that may be blended with the dominant lines are shown in light grey. Details of the dominant lines can be found in \autoref{tab:lines}.}
    \label{fig:nirspec_ID_3}
\end{figure}

Between 2.5$-$5.0\um, the NIR spectrum shows weak features, apart from two prominent features at around 2.9\um\ and 3.2\um, shown in \autoref{fig:nirspec_ID} and \autoref{fig:nirspec_ID_3}. The peak flux of the feature at 2.9\um\ is about half that of the 3.2\um\ feature and is dominated by two strong [Fe III] lines at 2.874 and 2.905\um. The red side of the peak may be blended with weaker [Ni II] 2.911\um\ and [Fe III] 3.044\um\ \citep{Flors2020}.

The broad, somewhat-boxy feature centered near 3.2\um\ is attributable to the [Ni I] 3.120\um\ line on the blue side and the [Fe III] 3.229\um\ line on the red side. Other potential blended contributors include [Fe III] 3.044\um, [Co II] 3.286\um\ and [Co II] 3.151\um\ lines. Models by \citet{Hoeflich2021} predict the [Ni I] 3.120\um\ line to be strong and narrow, with the [Fe III] 3.229\um\ line being weaker but broader. Our data suggest that the [Fe III] 3.229\um\ line is indeed broader, but more detailed modeling of all potential lines in this region is needed to unambiguously determine line strengths in this blended feature. We further discuss the line profile shapes and measurements of the 3.2\um\ feature in \autoref{sec:NIR_profile}. 

The remainder of the 2.5$-$5.0\um\ region shows only unidentifiable weak lines and strong, isolated lines do not reappear until the MIR around 6.5\um. Small bumps in the spectrum at $\sim$3.4\um\ (S/N $\simeq$ 9) and $\sim$3.8\um\ (S/N $\simeq$ 8) might be attributable to [Fe II] 3.393\um\ and [Ni III] 3.802\um, respectively, or a blend of unidentifiable features. Interestingly, several lines that are predicted to be strong do not appear in our \textit{JWST}/NIRSpec observations of SN~2021aefx. Models from \citet{Hoeflich2021} predict strong [Fe II] 4.076 and 4.115\um\ lines, and while there may be a small bump in the data in this region, it is weak (S/N $\simeq$ 7). These models also predict weaker lines of [Co I] 2.526\um\ and [Co I] 3.633\um\ that we do not see strongly in our data. However, the models from \citet{Hoeflich2021} were made for a subluminous SN, so it may be expected that the ionization state and relative line strengths are different between the models and our data. Additionally, the [Co III] 3.492\um\ line has been expected to contribute to the flux in the \textit{Spitzer} CH1 (3.6\um) photometric band \citep[e.g., see][]{Gerardy2007, Johansson2017}; however, it does not appear in our NIR spectrum. 

\subsection{MIR Emission Lines \label{sec:mir_lines}}

\begin{figure*}
    \centering
    \includegraphics[width=\textwidth]{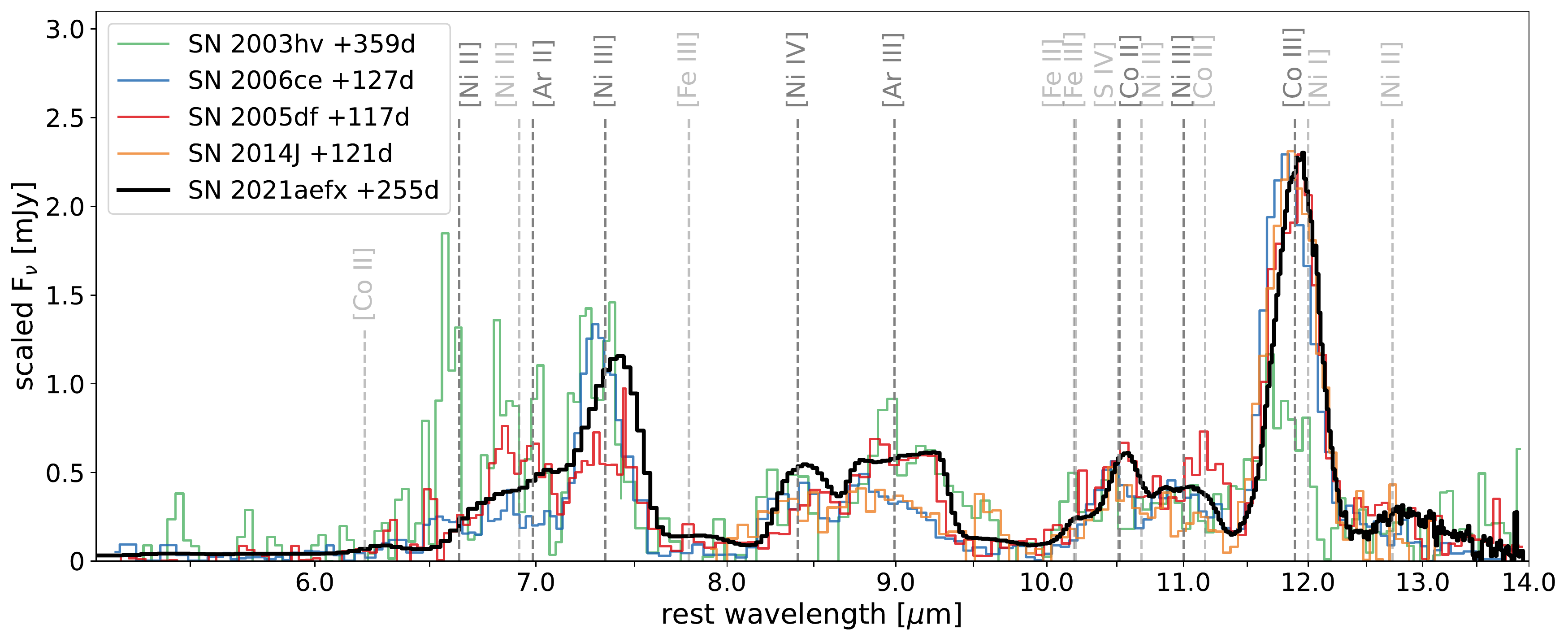}
    \caption{MIR line identifications for our \textit{JWST}/MIRI spectrum of SN~2021aefx (black) at $+$255d. We compare to scaled \textit{Spitzer}/IRS MIR spectra of SN~2005df (red) at $+$117d and SN~2003hv (green) at $+$359d \citep{Gerardy2007}, SN~2006ce (blue) $+$127d, and a MIR spectrum from GTC/CanariCam of SN~2014J (orange) at $+$121d \citep{Telesco2015}. All phases are given in rest-frame days relative to \textit{B}$_\text{max}$. The \textit{JWST}/MIRI spectrum shows impressive improvement in S/N, allowing for additional identification of weaker lines. The dominant emission lines are marked in grey, with uncertain line identifications marked in light grey. Details of the MIR lines can be found in \autoref{tab:lines}.}
    \label{fig:miri_ID}
\end{figure*}

Starting from the MIR line identifications by \citet{Gerardy2007} for SN~2005df, we mark the dominant emission lines in the \textit{JWST}/MIRI spectrum of SN~2021aefx in \autoref{fig:miri_ID}. Compared with the \textit{Spitzer}/IRS MIR spectra of SN~2005df and SN~2003hv ($+$117 and $+$359 rest-frame days from \textit{B}$_\text{max}$, respectively) \citep{Gerardy2007}, as well as an unpublished \textit{Spitzer} spectrum of SN~2006ce ($+$127 rest-frame days from \textit{B}$_\text{max}$; \citealt{Blackman2006}) (PID 30292, PI: P. Meikle) downloaded from the CASSIS database\footnote{\href{https://cassis.sirtf.com/}{https://cassis.sirtf.com/}} \citep{lebouteiller11}, and the Gran Telescopio Canarias (GTC)/CanariCam MIR spectrum of SN~2014J ($+$121 rest-frame days from \textit{B}$_\text{max}$)\citep{Telesco2015}, the improvement in signal-to-noise ratio of the \textit{JWST} spectrum is striking. Despite differences in phase, the MIR spectra all show fairly close agreement, with the same major emission lines present but varying in strength and shape. The \textit{JWST}/MIRI spectrum reveals additional, weaker emission lines and helps to confirm and clarify noisy lines in the \textit{Spitzer}/IRS and GTC/CanariCam spectra. Details of our MIR line identifications can be found in \autoref{tab:lines}. 

The optical and NIR nebular spectra are dominated by iron-group elements with most of the strongest lines attributed to Fe emission, while the most dominant emission lines in the MIR instead come from Co and Ni. Prominent emission lines from Ar, an intermediate-mass element, also emerge in the MIR, providing new physical insight into the SN emission structure. These MIR emission lines are significantly less blended than in the NIR, making their identification and subsequent fitting easier.

\subsubsection{Iron-Group Elements: Cobalt and Nickel \label{sec:CoNi_ID}}

The brightest MIR feature is the isolated [Co III] 11.888\um\ line, which shows close agreement with the SN~2005df and SN~2003hv spectra. The [Co II] 10.521\um\ line is also fairly strong and its peak is clearly resolved, although its base is blended with other lines on both sides. This line is potentially indistinguishably blended with the weaker and very nearby [S IV] 10.510\um\ line \citep{Gerardy2007}. [Co II] 10.521\um\ has roughly 25\% the strength of the [Co III] 11.888\um\ line, suggesting that there is comparatively little [Co II] emission.

On the blue side, the [Co II] 10.521\um\ line is blended with a previously unidentified line which creates a shoulder feature. We tentatively suggest that this unidentified line near 10.2\um\ is [Fe II] 10.189\um, [Fe III] 10.201\um, or a combination of both. This shoulder feature is also visible in SN~2003hv and potentially as a single pixel excess in SN~2005df, though it is not present in SN~2006ce or SN~2014J. It was not previously identified by \citet{Gerardy2007} due to high spectral noise.

The broad curve redward of the [Co II] 10.521\um\ line exhibits two distinct bumps, indicating blended emission. We identify the expected strong [Ni III] 11.002\um\ line and tentatively identify the weaker [Ni II] 10.682\um\ line as the main contributing species. Weak [Co II] 11.167\um\ emission may also broaden the redder of the two bumps, which is predominantly [Ni III] 11.002\um. Inspection of the SN~2005df spectrum reveals a single pixel excess at the location of the bluer bump, the SN~2003hv spectrum also shows a peak (though without the \textit{JWST}/MIRI spectrum for comparison, these very faint signals look like, and could be, noise). SN~2014J clearly exhibits two distinct peaks in this region while the SN~2006ce spectrum looks smoothly blended into one peak. All of the supernovae exhibit emission consistent with the [Ni III] 11.002\um\ line.

The second brightest MIR feature, at roughly 50\% of the [Co III] 11.888\um\ line strength, is the relatively isolated [Ni III] 7.349\um\ line. This feature in SN~2021aefx is slightly weaker in relative strength than in SN~2006ce and SN~2003hv, but is stronger than in SN~2005df.

The [Ni IV] 8.405\um\ line in SN~2021aefx has a roughly equal peak flux as the neighboring [Ar III] 8.991\um\ line and is fairly well isolated, with minimal blending. This line was identified for SN~2005df, though its strength was only about 30\% the strength of the [Ar III] 8.991\um\ line. SN~2006ce and SN~2014J also exhibit the [Ni IV] 8.405\um\ line, and like SN~2021aefx, it has roughly equal peak flux to the [Ar III] 8.991\um\ line. The [Ni IV] feature in SN~2003hv was only speculatively identified by \citet{Gerardy2007} due to noise, but it is confirmed more clearly when compared to the MIRI spectrum of SN~2021aefx.

While \citet{Gerardy2007} found no convincing detection of the [Ni II] 6.636\um\ line (partially due to excess noise in the 7.4\um\ region from the overlapping edges of spectral orders), our \textit{JWST}/MIRI spectrum displays a clear ``shoulder'' feature on the blue side of the 7$-$8\um\ region explained by [Ni II] 6.636\um\ blending with the nearby [Ar II] 6.985\um\ line. 

Further analysis of these iron-group element features in the MIR spectrum is given in \autoref{sec:Co_profile} and \autoref{sec:Ni_profile}, where we fit simple geometric line profiles and estimate kinematic properties for each identified line.

\subsubsection{Intermediate-Mass Elements: Sulfur and Argon \label{sec:Ar_ID}}

[S IV] 10.510\um\ line emission theoretically contributes to the [Co II] 10.521\um\ peak \citep{Gerardy2007}; however, it is too closely blended and our spectral resolution is too low for direct identification or further analysis, especially given the additional blending with Ni on both sides. Detailed theoretical models will be needed to disentangle the contribution of [S IV] 10.510\um.

Two important argon emission lines appear in the MIR. In the MIRI spectrum, the [Ar III] 8.991\um\ line is only slightly blended with the [Ni IV] 8.405\um\ line. We see a broad, well-resolved flat-topped profile indicative of a lack of emission at low projected velocities implying a spherical shell of emission. This boxy [Ar III] shape differs significantly from the forked [Ar III] and [Ar II] profiles that \citet{Gerardy2007} found for both SN~2005df and SN~2003hv, attributed to an asymmetric ring of emission. SN~2006ce and SN~2014J appear to have an asymmetrically sloped [Ar III] profile, highlighting interesting differences in the distribution of argon between these supernovae. 

Visible as a small bump on top of the broad blue wing of the [Ni III] 7.349\um\ feature, the [Ar II] 6.985\um\ line in SN~2021aefx is nearly completely blended into the Ni emission lines that surround it. With such strong blending, it is difficult to conclusively determine the shape of this line; we further analyze the shape of the [Ar III] 8.991\um\ and [Ar II] 8.405\um\ lines and discuss their implications in \autoref{sec:Ar_profile}.

\subsubsection{Unidentified and Speculative Features \label{sec:un_ID}}

\citet{Gerardy2007} suggest possible detection of silicon monoxide (SiO) molecular emission in the $\sim$7.5-8\um\ region of SN~2005df, corresponding to the fundamental ($\Delta_v = 1$) rovibrational band. \citet{Hoeflich1995} concluded that CO and SiO might form in subluminous SNe Ia with very low $^{56}$Ni yield, and SN~2005df was a subluminous SN Ia, making detection of SiO an interesting possibility. However, SN~2021aefx is a normal SN Ia; furthermore, our full NIR $+$ MIR spectrum does not show convincing evidence for CO or SiO fundamental emission elsewhere in the spectrum. Thus, we favor an explanation for the weak emission feature at $\sim$7.8\um\ by [Fe III] 7.791\um, which is predicted by models from \citet{Flors2020}.

We speculate that the faint emission feature at $\sim$6.2\um\ is [Co II] 6.214\um. [Ni II] 12.729\um\ may be detected redward of the strong [Co III] 11.888\um\ line, but an increase in noise toward the end of the spectrum prevents conclusive detection. Finally, a small spike on top of the red side of the [Co III] 11.888\um\ peak is tentative evidence for [Ni I] 12.001\um.

\begin{deluxetable*}{llccrrcccc}[b!]
\tablecaption{Line Identification and Fitting for SN~2021aefx\tablenotemark{a} \label{tab:lines}}
\tablehead{\colhead{$\lambda_{\textrm{rest}}$} & \colhead{Species} & \colhead{Fit Profile} & \colhead{$\lambda_{\text{peak}}$} & \colhead{$v_\text{peak}$} & \colhead{FWHM} & Transition & $A_{ki}$ & $E_{lu}$\\
($\mu$m) & & & ($\mu$m) & (km s$^{-1}$) & (km s$^{-1}$) & & &  (eV)}
\tablecolumns{6}
\startdata
\multicolumn{6}{l}{\textbf{Optical + NIR Lines}}\\ 
0.589 & [Co III]  & Gaussian & 0.591 & 1000 $\pm$ \hphantom{1}500  & 10,800 $\pm$ 2500 & a$^4$F$_{9/2}-$a$^2$G$_{9/2}$ & \num{4.20e-1} & 0.000$-$2.105\\
0.716 & [Fe II]  & Gaussian & 0.717 & 600 $\pm$ \hphantom{1}100  & 8900 $\pm$ 2200 & a$^4$F$_{9/2}-$a$^2$G$_{9/2}$ & \num{1.46e-1} & 0.232$-$1.964\\
0.738 & [Ni II]  & Gaussian & 0.740 & 900 $\pm$ \hphantom{1}200  & 9700 $\pm$ 3600 &  $^2$D$_{5/2}-^2$F$_{7/2}$ & \num{2.30e-1} & 0.000$-$1.680\\
1.257 & [Fe II]   & Gaussian & 1.263 & 1300 $\pm$ 1500 & 9400 $\pm$ 1500 & a$^6$D$_{9/2}-$a$^4$D$_{7/2}$ & \num{4.74e-03} & 0.000$-$0.986\\
1.547 & [Co II]  & Gaussian & 1.545 & $-$400 $\pm$ 1300 & jussi6100 $\pm$ 2200 & a$^5$F$_5-$b$^3$F$_4$  &  \num{2.81e-02}  &  0.415$-$1.217\\
1.644 & [Fe II]   & Gaussian & 1.651 & 1300 $\pm$ 1200 & 11,100 $\pm$ 2400 & a$^4$F$_{9/2}-$a$^4$D$_{7/2}$ &  \num{6.00e-03} & 0.232$-$0.986\\
1.939 & [Ni II]   & Gaussian & 1.949 & 300 $\pm$ 1000 & 11,300 $\pm$ 1000 & $^4$F$_{9/2}-^2$F$_{7/2}$ &  \num{8.70e-02} & 1.041$-$1.680\\
2.219 & [Fe III]  & Gaussian & 2.226 & 1100 $\pm$ \hphantom{1}800 & 9900 $\pm$ \hphantom{1}800 & $^3$H$_6-^3$G$_5$ & \num{3.40e-02} & 2.486$-$3.045\\
2.874 & [Fe III]  & \multirow{2}{*}{Gaussian} & 2.867 & \multirow{2}{*}{$-$100 $\pm$ \hphantom{1}500} & \multirow{2}{*}{13,900 $\pm$ 4300} & $^3$F4$_4-^3$G$_4$ & \num{2.70e-02} & 2.661$-$3.092 \\
2.905 & [Fe III]  & & 2.906 & & & $^3$F4$_3-^3$G$_3$ & \num{2.90e-02} & 2.690$-$3.117 \\
3.120  & [Ni I]   & Gaussian & 3.118 & $-$200 $\pm$ \hphantom{1}400 & 6300 $\pm$ 1900 & $^3$D$_3-^1$D$_2$ & \num{7.80e-02} & 0.025$-$0.423\\
3.229  & [Fe III] & Gaussian & 3.209 & $-$1900 $\pm$ \hphantom{1}400 & 11,300 $\pm$ 3000 &  $^3$F4$_4-^3$G$_5$ &  \num{1.70e-02} &  2.661$-$3.045\\
\\
\multicolumn{6}{l}{\textbf{MIR Lines}} \\ 
6.636  & [Ni II]   & Gaussian & 6.725  & 4000 $\pm$ 1500 & 12,000 $\pm$ 3500 & $^2$D$_{5/2}-^2$D$_{3/2}$ & \num{5.54e-02} & 0.000$-$0.187\\
6.985  & [Ar II]   & Gaussian & 7.084  & 4000 $\pm$ 1500 & 20,500 $\pm$ 4100 & $^2$P$^o_{3/2}-^2$P$^o_{1/2}$ & \num{4.23e-02} & 0.000$-$0.177\\
7.349  & [Ni III]  & Gaussian & 7.422  & 3000 $\pm$ 1400 & 11,200 $\pm$ 1300 & $^3$F$_4-^3$F$_3$ & \num{6.50e-02} & 0.000$-$0.169\\
8.405  & [Ni IV]   & Sphere   & 8.447  & 1300 $\pm$ 1200 & 13,600 $\pm$ \hphantom{1}600 & $^4$F$_{9/2}-^4$F$_{7/2}$ & \num{5.70e-02} &  0.000$-$ 0.148\\
8.991  & [Ar III]  & Shell    & 9.012  & 700  $\pm$ 1100 & 23,700 $\pm$ \hphantom{1}600 & $^3$P$_2-^3$P$_1$ & \num{3.10e-02} & 0.000$-$0.138 \\
10.521 & [Co II]   & Gaussian & 10.562 & 1200 $\pm$ 1000 & 8300 $\pm$ 1300 & a$^3$F$_4-$a$^3$F$_3$ &  \num{2.24e-02} & 0.000$-$0.118\\
11.002 & [Ni III]  & Gaussian & 11.051 & 1300 $\pm$ \hphantom{1}900  & 10,700 $\pm$ 2400 &  $^3$F$_3-^3$F$_2$ &  \num{2.70e-02} & 0.169$-$ 0.281\\
11.888 & [Co III]  & Gaussian & 11.911 & 500 $\pm$ \hphantom{1}900 & 10,200 $\pm$ 1300 & a$^4$F$_{9/2}-$a$^4$F$_{7/2}$ & \num{2.01e-02} & 0.000$-$0.104\\
\\
\multicolumn{6}{l}{\textbf{Tentative Lines}}\\
2.911  & [Ni II]?  & Gaussian & 2.913 & 4800 $\pm$ 2000 & $-$400 $\pm$ 2600 & $^4$F$_{5/2}-^2$F$_{7/2}$ & \num{1.40e-02} & 1.254$-$1.680\\
3.044  & [Fe III]? & -- & -- & --\hphantom{abcde} & --\hphantom{abcde} & $^3$F4$_2-^3$G$_3$ &  \num{1.80e-02} & 2.710$-$3.117\\
3.286  & [Co II]?  & Gaussian & 3.287 & $-$300 $\pm$ \hphantom{1}400 & 5900 $\pm$ 1200 & c$^3$F$_4-$a$^1$F$_3$ & \num{1.86e-03} & 5.089$-$5.467\\
6.214  & [Co II]?  & -- & -- & --\hphantom{abcde} & --\hphantom{abcde} & a$^1$D$_2-$a$^3$P$_2$ & \num{3.08e-02} & 1.445$-$1.644 \\
6.920  & [Ni II]?  & Gaussian & 7.055 & 5800 $\pm$ 2000 & 12,000 $\pm$ 4000 & $^2$F$_{7/2}-^2$F$_{5/2}$ & \num{4.71e-02} & 1.680$-$1.859\\
6.985  & [Ar II]?  & Shell    & 7.039 & 2300 $\pm$ 1500 & 23,700 $\pm$ \hphantom{1}600 & $^2$P$^o_{3/2}-^2$P$^o_{1/2}$ & \num{4.23e-02} & 0.000$-$0.177\\
7.791  & [Fe III]? & -- & -- & --\hphantom{abcde} & --\hphantom{abcde} & $^3$P4$_2-^3$P4$_1$ & \num{4.70e-02} & 2.406$-$2.565\\
10.189  & [Fe II]?  & \multirow{2}{*}{Gaussian} & \multirow{2}{*}{10.235} & \multirow{2}{*}{1000 $\pm$ 1000} & \multirow{2}{*}{6700 $\pm$ 1700} & b$^4$P$_{5/2}-$b$^4$P$_{3/2}$ & \num{2.30e-02} & 2.583$-$2.704\\
10.201 & [Fe III]?  &  &  &  &  &  $^3$H$_4-^3$F4$_4$ &  \num{1.60e-03} & 2.539$-$2.661\\
10.510 & [S IV]?  & -- & -- & --\hphantom{abcde} & --\hphantom{abcde} & $^2$P$^o_{1/2}-^2$P$^o_{3/2}$ & \num{7.30e-03} & 0.000$-$0.118\\
10.682 & [Ni II]?  & Gaussian & 10.842 & 4500 $\pm$ 1000  & 4200 $\pm$ 1500 & $^4$F$_{9/2}-^4$F$_{7/2}$ &  \num{2.71e-02} & 1.041$-$1.157\\
11.167 & [Co II]?   & Gaussian & 11.214 & 1300 $\pm$ \hphantom{1}900  & 3100 $\pm$ 1400 & b$^3$F$_4-$b$^3$F$_3$ & \num{1.88e-02} & 1.217$-$1.328\\
12.002 & [Ni I]? & -- & -- & $\sim$ 1500\hphantom{abc} & --\hphantom{abcde} & $^3$D$_2-^3$D$_1$ &  \num{2.10e-02} & 0.109$-$0.212\\
12.729 & [Ni II]? & -- & -- & --\hphantom{abcde} & --\hphantom{abcde} & $^4$F$_{7/2}-^4$F$_{5/2}$ & \num{2.76e-02} & 1.157$-$1.254
\enddata
\tablenotetext{a}{Line information from the Atomic Line List version 3.00b4}
\end{deluxetable*}

\section{Line Velocities and Profiles \label{sec:profiles}}

At late times in the nebular phase, the supernova ejecta opacity drops and emission streams freely from all regions, revealing important properties of the ejecta composition and ionization structure. The shape of the nebular emission lines is determined by the ejecta emissivity, which depends on both the density and excitation \citep[for a review, see][]{jerkstrand_spectra_2017}. Several simple ejecta geometries that produce common line profiles include a uniform sphere resulting in a parabolic shape, a uniform spherical thick shell resulting in a boxy, flat-topped shape with parabolic wings, and a Gaussian density sphere resulting in a Gaussian line profile \citep{jerkstrand_spectra_2017}. The MIR, where the features are comparatively isolated, is particularly useful for inferring the kinematic distribution of the emission. 

Most of the lines that we identify can be well modelled by a superposition of Gaussian line profiles. Using the $+$266d emission line model of SN~2015F from \citet{Flors2020}, and following the approach of \citet{Maguire2018} and \citet{Flors2018, Flors2020}, we model the superposition of all [Fe II], [Fe III], [Ni II], [Co II], and [Co III] lines contributing to selected optical and NIR features. The relative line strengths of each line are fixed by the model, and all lines of the same species are restricted to have the same Gaussian width and kinematic offset from the central wavelength. Because this model was computed for temperatures and densities specific to SN~2015F, not SN~2021aefx, we do not attempt to fit the entire optical or NIR spectrum, but rather fit the model to selected regions containing features of interest.

Past 3\um, the emergence of lines from species not included in the SN~2015F model from \citet{Flors2020} prevents us from modeling each feature in such a thorough and self-consistent way. Furthermore, the model relative line strengths begin to deviate significantly from the data for the NIR 3.2\um\ feature and the MIR. Thus, we fit each feature redward of 3\um\ as a superposition of basic geometric line-emission profiles for each distinguishable contributing line, allowing the amplitude, kinematic offset, and width of each individual line to be free parameters. Fit profile shapes for each line are chosen based upon visual inspection of best overall feature fit. A proper, bespoke model for SN~2021aefx is beyond the scope of this paper, but will be the focus of future effort.

We use UltraNest \citep{Buchner2021}, a Bayesian inference package for parameter estimation using nested sampling, to fit our line profiles and recover uncertainties on our measurements of kinematic offset ($v_\text{peak}$) and full-width half-maximum (FWHM). The likelihood function optimized in the UltraNest fitting is given by:
\begin{equation*}
    \ln(\text{likelihood}) = -\frac{1}{2} \sum \left[\frac{\left(\text{data}-\text{model}\right)^2}{s^2} + \ln(2 \pi s^2)\right]
\end{equation*}
where
\begin{equation*}
    s^2 = \text{data\_uncertainty}^2 + f^2~\text{model}^2
\end{equation*}
and the uncertainties are underestimated by some fractional amount $f$ that is marginalized over in the fit. For the MIR lines, we include a systematic uncertainty in the wavelength calibration of 0.034\um, derived from the root-mean-square of the wavelength calibration residuals in the region encompassing the SN features (6.5--12.5\um), and add this in quadrature to the kinematic offset uncertainties. The typically low resolution of our data results in instrumental broadening ($c \Delta\lambda/\lambda$) that is significant: $\sim$4500\kms\ at 1.25\um\ and $\sim$1000\kms\ at 3.3\um\ for the NIR, and $\sim$1900\kms\ at 6.5\um\ and $\sim$450\kms\ at 12\um\ for the MIR. We remove this in quadrature from our FWHM measurements and impose a floor on our line centroid uncertainty equal to one-third the instrument resolution. For a sense of scale, the instrumental resolution is roughly 50\% of the FWHM of the [Co III] 1.257\um\ line, 10\% of the FWHM of the [Fe III] 3.229\um\ line, 15\% of the FWHM of the [Ni III] 7.349\um\ line, and 5\% of the FWHM of the [Co III] 11.889\um\ line.
 
Our line fits do not include any radiative transfer and we only attempt basic accounting of line blending by superposition. \autoref{tab:lines} gives the chosen fit profile shape, measured peak wavelength, kinematic offset, FWHM, and the estimated uncertainties of each fitted feature. 

\subsection{Optical $+$ NIR: Iron and Nickel}

\begin{figure}
    \centering
    \includegraphics[width=0.5\textwidth]{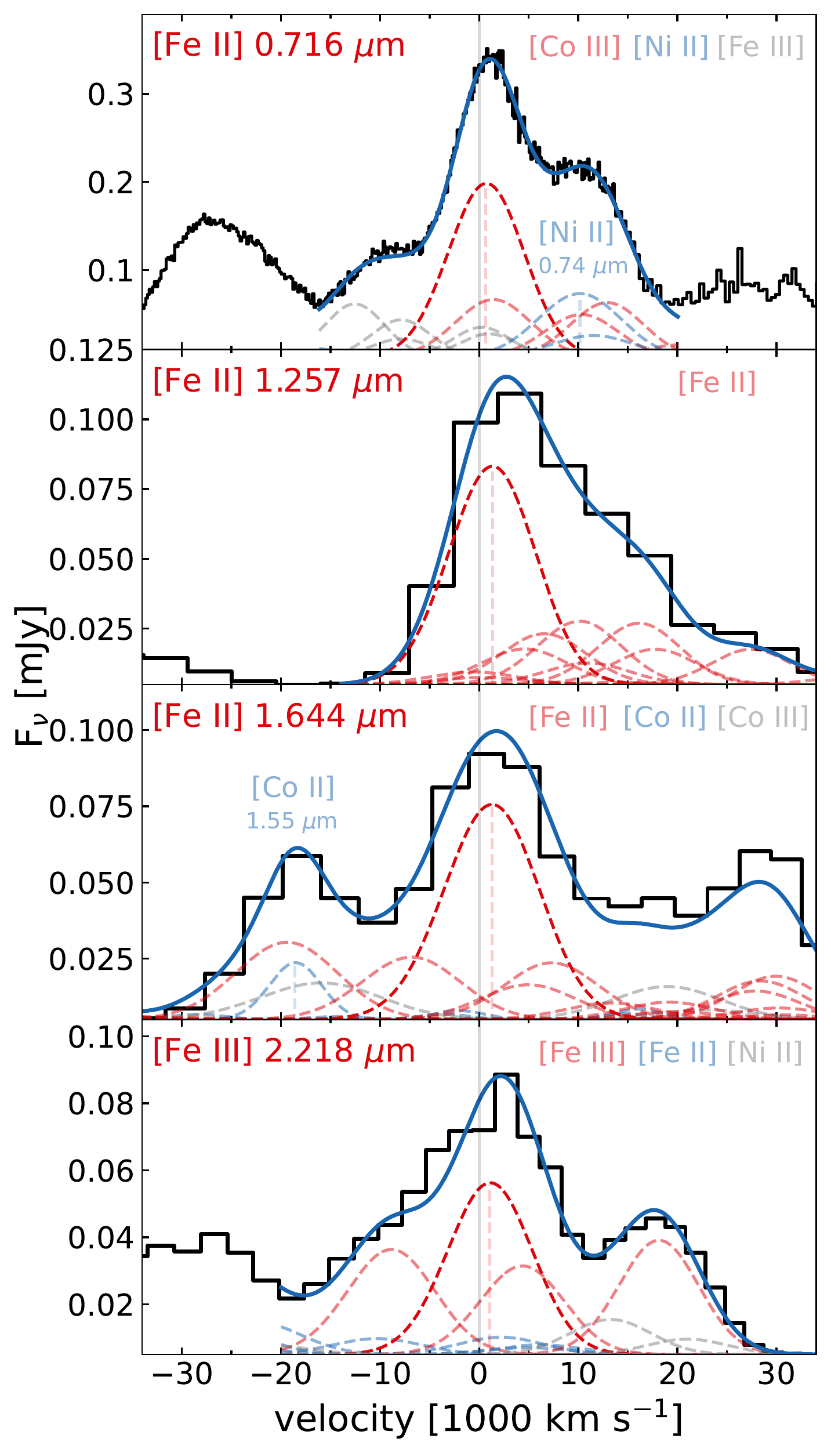}
    \caption{Optical $+$ NIR emission-line profiles of SN~2021aefx at $+$255d (black) for complex spectral features dominated by iron. The model profile for the central line is shown in dashed red, with contributions from other lines shown in dashed light red and dashed light blue, and the full modeled superposition of lines shown in solid blue.}
    \label{fig:nir_vels_2}
\end{figure}

Using the SN~2015F model from \citet{Flors2020} as described above, we fit the NIR [Fe II] 1.257\um, [Fe II] 1.644\um, and [Fe III] 2.218\um\ features, shown in \autoref{fig:nir_vels_2}. We find agreement within the uncertainties, in both $v_\text{peak}$ and FWHM, between all of these lines, which exhibit a moderately redshifted $v_\text{peak}$. 

Like the models by \citet{Maguire2018}, the SN~2015F model from \cite{Flors2020} only contains significant line contributions in the 1.3\um\ region from [Fe II], as shown in \autoref{fig:nir_vels_2}. The model fits the data well in this region and the measured FWHM and $v_\text{peak}$ agree closely with those measured from the optical [Fe II] 0.716\um\ and NIR [Fe II] 1.644 \um\ lines. This supports the conclusion of \citet{Maguire2018} that the 1.3\um\ feature is a relatively contaminant free way to measure line velocities and widths for [Fe II].

We also fit the optical 7300~\AA\ line complex of [Fe II] 0.716\um\ and [Ni II] 0.738\um\ with the SN~2015F model, as shown in \autoref{fig:nir_vels_2}. The measured FWHM and $v_\text{peak}$ for the [Fe II] 0.716\um\ line are within the uncertainties of the NIR Fe lines. The [Ni II] 0.738\um\ line is consistent in FWHM with the other [Ni II] and [Ni III] lines in the NIR and MIR. Its $v_\text{peak}$ is comparable to [Ni II] 1.939\um, [Ni III] 11.002\um, and [Ni IV] 8.405\um, but lower than [Ni II] 6.644\um\ and [Ni III] 7.349\um, which may be partly attributed to the imperfect \textit{JWST} MIRI/LRS wavelength calibration that is worse at the shorter wavelengths. Overall, from the optical 7300~\AA\ line complex, the [Fe II] 0.716\um\ and [Ni II] 0.738\um\ kinematics are roughly consistent with the Fe and Ni kinematics from the NIR and MIR.

We detect the [Ni II] 1.939\um\ line feature in our NIR spectrum, shown in \autoref{fig:ni_vels}, which has also been seen in other ground-based studies of the NIR \citep{Dhawan2018, Flors2020, Blondin2022}. Fitting this feature with the 2015F model, we find a FWHM that is comparable to the FWHM of Ni II in the MIR. The $v_\text{peak}$ is within the uncertainties of the [Ni II] 0.738\um, [Ni III] 7.349\um, [Ni III] 11.00\um, and [Ni IV] 8.41\um\ lines and slightly smaller than the [Ni II] 6.64\um\ $v_\text{peak}$, again possibly affected by the MIR wavelength calibration.

Following \citet{Flors2020}, we find that while we cannot rule out weak [Ca II] line contamination in the 7300~\AA\ line complex, strong [Ca II] contamination in the optical would require a weaker [Ni II] 0.716\um\ line, thus predicting a weaker [Ni II] 1.939\um\ line since no [Ca II] is present there. Our fits to [Ni II] 0.738\um\ and [Ni II] 1.939\um\ have similar amplitudes and we do not see a weaker [Ni II] 1.939\um\ line than expected from [Ni II] 0.738\um. Thus, in agreement with the findings by \citet{Maguire2018} and \citet{Flors2020}, we do not find a compelling reason to invoke contamination from [Ca II] to reconcile the differences in kinematic properties between the optical and NIR $+$ MIR.

\begin{figure}
    \centering
    \includegraphics[width=0.5\textwidth]{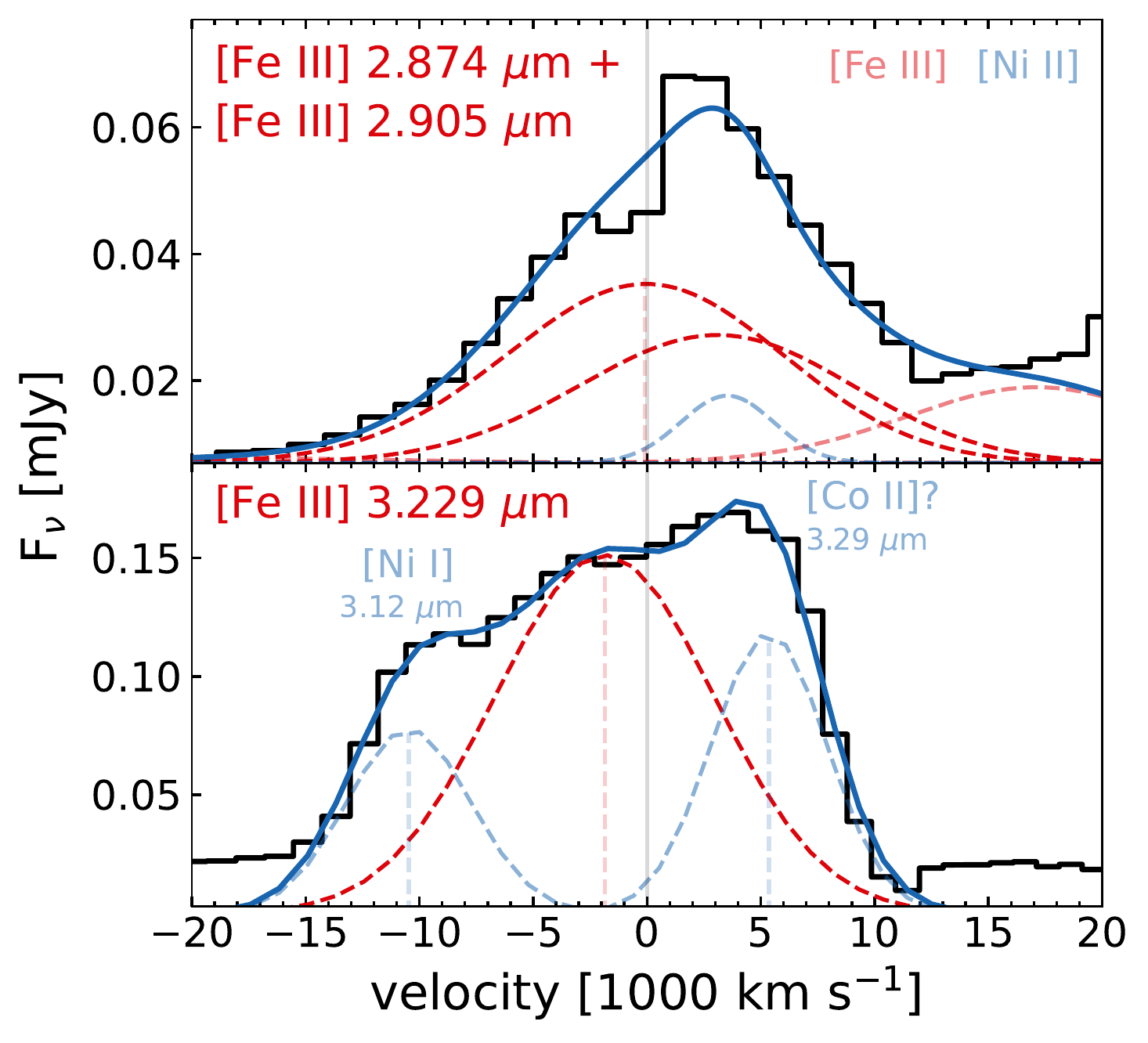}
    \caption{NIR emission-line profiles of SN~2021aefx at $+$255d (black) for the 3\um\ region features. The model profile for the central line is shown in dashed red, with contributions from other lines shown in dashed blue and the full modeled superposition of lines shown in solid blue.}
    \label{fig:nir_vels}
\end{figure}

\subsubsection{NIR: 2.9 and 3.2\um\ Features \label{sec:NIR_profile}}

The [Fe III] 2.874\um\ $+$ [Fe III] 2.905\um\ line feature, fit to the SN~2015F model from \citet{Flors2020}, has contributions from [Ni II] 2.911\um\ and [Fe III] 3.044\um, shown in \autoref{fig:nir_vels}. This feature exhibits a broad FWHM consistent with [Fe III] 2.218\um\ and [Fe III] 3.229\um, and a blueshifted $v_\text{peak}$ between the other two NIR [Fe III] lines.

We fit the feature at 3.2\um\ with a superposition of three Gaussian emission distribution profiles, shown in \autoref{fig:nir_vels}. The strongest lines expected in this feature are [Ni I] 3.120\um\ and [Fe III] 3.229\um, but to improve the fit of the full feature, we include a third [Co II] 3.286\um\ line. This fit produces a strong, broad [Fe III] 3.229\um\ line consistent in FWHM with [Fe III] 2.874\um\ and [Fe III] 2.218\um, but with a large blueshifted $v_\text{peak}$. This may indicate that a more comprehensive model is needed to explain the 3.2\um\ feature. The fits to the [Ni I] 3.120\um\ and [Co II] 3.286\um\ lines show weaker, narrower profiles of similar strength, FWHM, and low kinematic offset.

Alternatively, we can fit the 3.2\um\ feature nearly equally well without [Co II] 3.286\um, but it requires boxy, flat-topped profiles for both [Ni I] 3.120\um\ and [Fe III] 3.229\um. In this alternative fit, both lines are of similar strengths, nearly equally broad, and display large redshifts. However, these profiles would imply a large central hole of both [Fe III] and [Ni I] emission that would be difficult to explain physically. The Fe comes from $^{56}$Ni decay and is expected to be broad, whereas the observed Ni must be stable $^{58}$Ni, as all radioactive $^{56}$Ni has decayed away at this phase. The stable Ni is expected to be centrally concentrated and thus with a narrower, peaked profile. Furthermore, flat-topped profiles are not seen in other Ni and Fe features throughout the optical, NIR, and MIR spectra. Therefore, we favor the Gaussian profile fits with inclusion of an unexpectedly strong [Co II] 3.286\um\ line. More detailed future modeling of this feature may reveal additional contributing lines.

\begin{figure}
    \centering
    \includegraphics[width=0.5\textwidth]{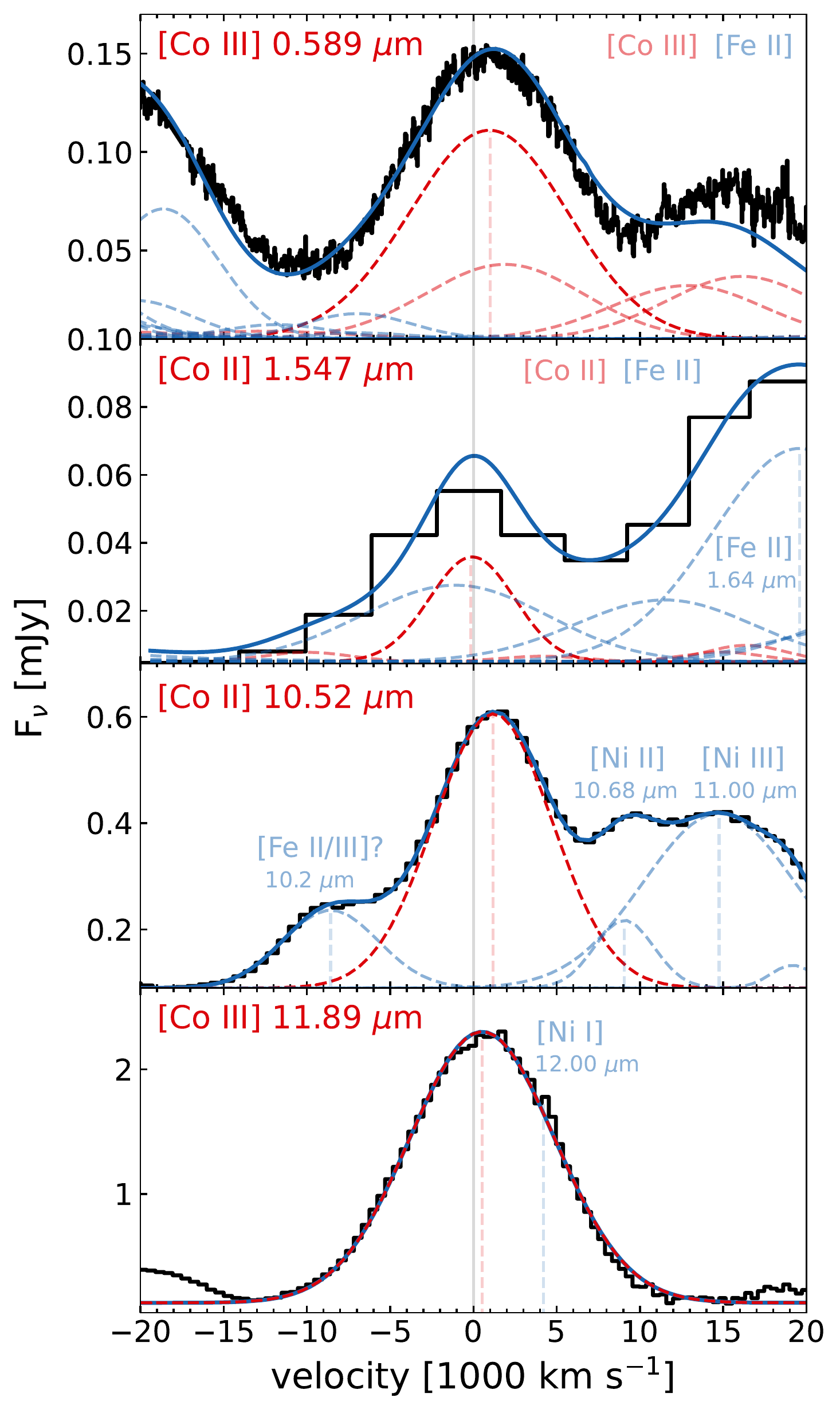}
    \caption{Cobalt emission-line profiles of SN~2021aefx at $+$255d (black) across the optical, NIR, and MIR compared with model line emission from Gaussian distributions of emission. The model profile for the central line is shown in dashed red, with contributions from other lines shown in dashed blue and the full modeled superposition of lines shown in solid blue.}
    \label{fig:co_vels}
\end{figure}

\subsection{MIR: Cobalt \label{sec:Co_profile}}

Shown in \autoref{fig:co_vels}, the [Co III] 11.888\um\ line is well fit by a fairly broad Gaussian profile and low kinematic offset from the host-galaxy rest frame. \citet{Gerardy2007} find evidence for a parabolic, slightly flat-topped [Co III] 11.888\um\ profile for SN~2005df resulting from a spherical distribution with a hollow inner region, and a significantly blueshifted, fairly flat and weak [Co III] 11.888\um\ line for SN~2003hv. SN~2021aefx shows a Gaussian [Co III] distribution with clear wings not expected from a uniform spherical distribution. However, close inspection of the Gaussian fit shows that the peak may be marginally flat topped, potentially indicating a Gaussian distribution with a small central hole corresponding to the electron capture zone where little Co is produced \citep{Gerardy2007}.

The [Co II] 10.521\um\ line is blended so closely with the predicted weaker [S IV] 10.510\um\ line that they are indistinguishable and we model them as one line. We model the full blended $\sim$10-11.3\um\ feature as a linear combination of five Gaussians: [Fe II] 10.189\um/[Fe III] 10.201\um, [Co II] 10.521\um, [Ni II] 10.682\um, [Ni III] 11.002\um, and [Co II] 11.167\um. The FWHM, $v_\text{peak}$, and Gaussian profile shape agree nicely between the [Co III] 11.888\um\ and [Co II] 10.521\um\ lines.

For comparison to the optical and NIR, we also fit the SN~2015F model from \citet{Flors2020} to the [Co III] 0.589\um\ and [Co III] 1.547\um\ lines in \autoref{fig:co_vels}. The [Co II] and [Co III] measurements agree quite well across the optical, NIR, and MIR and we conclude that the emission structure between Co ionization states is similar, though the MIR line strengths suggest there is less emission from [Co II]. The tentative identifications of [Co II] 11.167\um\ and [Co II] 3.286\um\ show decent agreement as well, though they are significantly less broad and slightly blueshifted, respectively.

\subsection{MIR: Nickel \label{sec:Ni_profile}}

\begin{figure}
    \centering
    \includegraphics[width=0.47\textwidth]{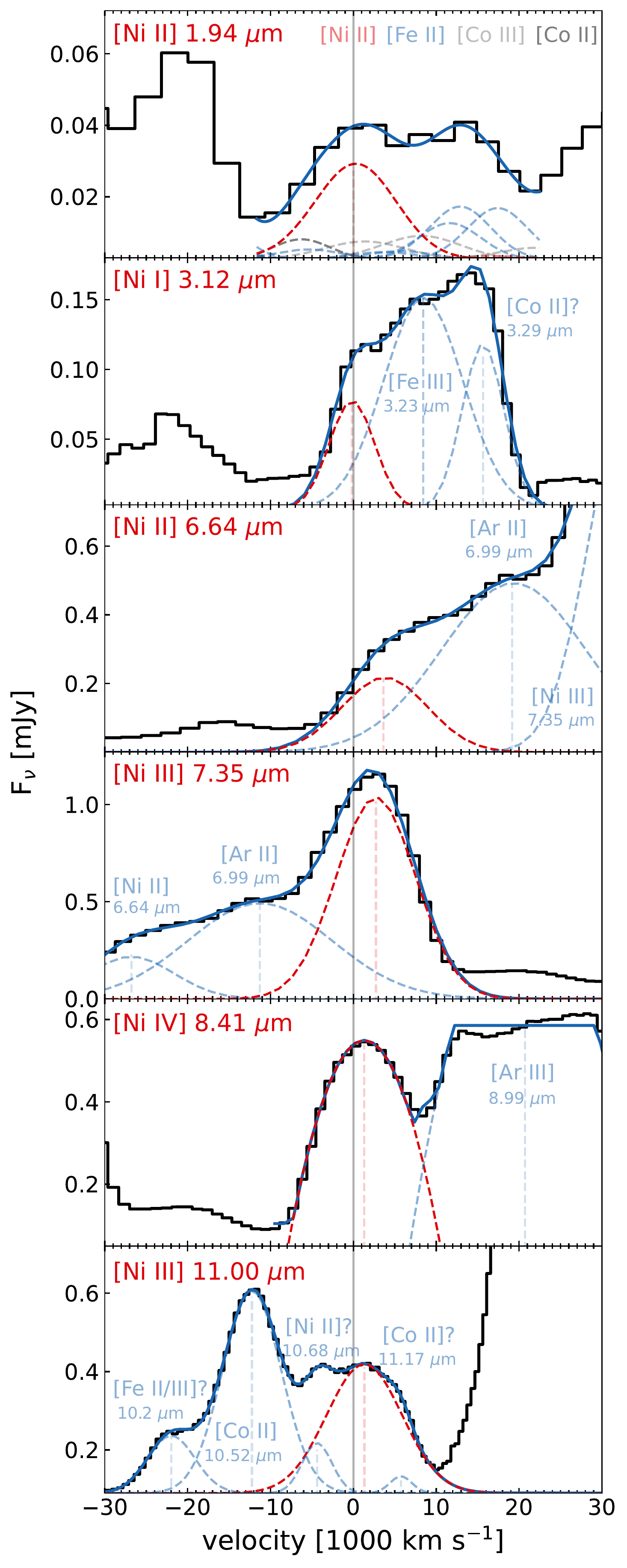}
    \caption{Observed NIR and MIR nickel emission-line profiles for SN~2021aefx at $+$255d (black) compared to model line emission from Gaussian distributions, except for [Ni IV] 8.405\um\ which is better fit by emission from a uniform spherical emission distribution. The model profile for the central line is shown in dashed red, with contributions from other lines shown in dashed blue and the full modeled superposition of lines shown in solid blue.}
    \label{fig:ni_vels}
\end{figure}

The SN~2021aefx MIR nickel emission-line profiles, shown in \autoref{fig:ni_vels}, are all blended with other emission-line profiles, with the [Ni IV] 8.405\um\ and [Ni III] 7.349\um\ lines being the most isolated. Despite blending, the \textit{JWST}/MIRI Ni lines in SN~2021aefx are significantly better resolved than the \textit{Spitzer}/IRS and GTC/CanariCam spectra and we can fit the blended Ni lines well by Gaussian profiles, except for [Ni IV] 8.405\um\ which prefers a parabolic profile. The wings of a Gaussian profile for [Ni IV] contribute too much to the boxy [Ar III] 8.991\um\ feature, implying the [Ni IV] emission arises from a uniform spherical, rather than Gaussian, geometry.

The Ni lines show significantly redshifted kinematic offsets, except for neutral [Ni I] 3.120\um, which is also the only NIR Ni line we fit. [Ni II] 6.636\um\ and [Ni III] 7.349\um\ exhibit the highest $v_\text{peak}$ values, which may be partially due to the MIRI wavelength solution being more uncertain at shorter wavelengths. Taking the wavelength uncertainties into account, these lines are within the uncertainty range of the [Ni IV] 8.405\um\ and [Ni III] 11.002\um\ measurements. These MIR [Ni II] and [Ni III] lines are consistent in FWHM, while the NIR [Ni I] 3.120\um\ line is narrower. The tentatively identified [Ni II] 10.682\um\ line shows a highly redshifted $v_\text{peak}$, consistent with [Ni II] 6.636\um\ and [Ni III] 7.349\um, but with an inconsistently narrower FWHM.

\subsection{MIR: Argon \label{sec:Ar_profile}}

\begin{figure}[t]
    \centering
    \includegraphics[width=0.5\textwidth]{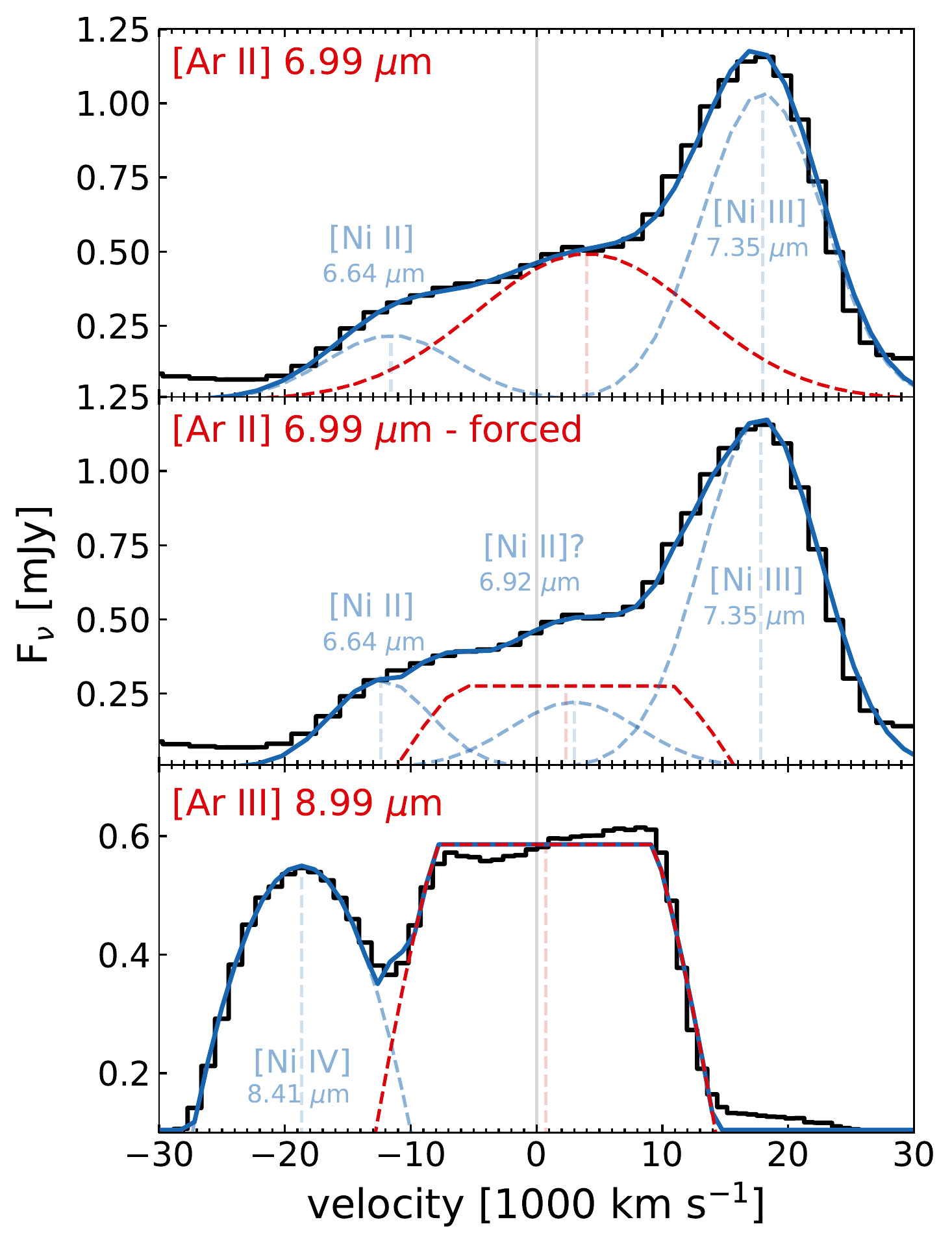}
    \caption{Observed argon emission-line profiles for SN~2021aefx at $+$255d (black) compared to model line emission from a Gaussian emission distribution for [Ar II] 6.985\um\ (top); a thick uniform shell/hollow uniform sphere emission distribution for [Ar III] 8.991\um\ with an inner radius corresponding to $v_\text{min} =$ 8700 $\pm$ 200\kms\ and an outer radius corresponding to $v_\text{max} =$ 13,500 $\pm$ 300\kms\ (bottom); and a thick shell distribution for [Ar II] 6.985\um\ which has been forced to have the same inner and outer radius as the [Ar III] fit (middle). The model profile for the central line is shown in dotted red, with contributions from other lines shown in dotted blue and the full modeled superposition of lines shown in solid blue.}
    \label{fig:ar_vels}
\end{figure}

\begin{figure*}
    \centering
    \includegraphics[width=0.9\textwidth]{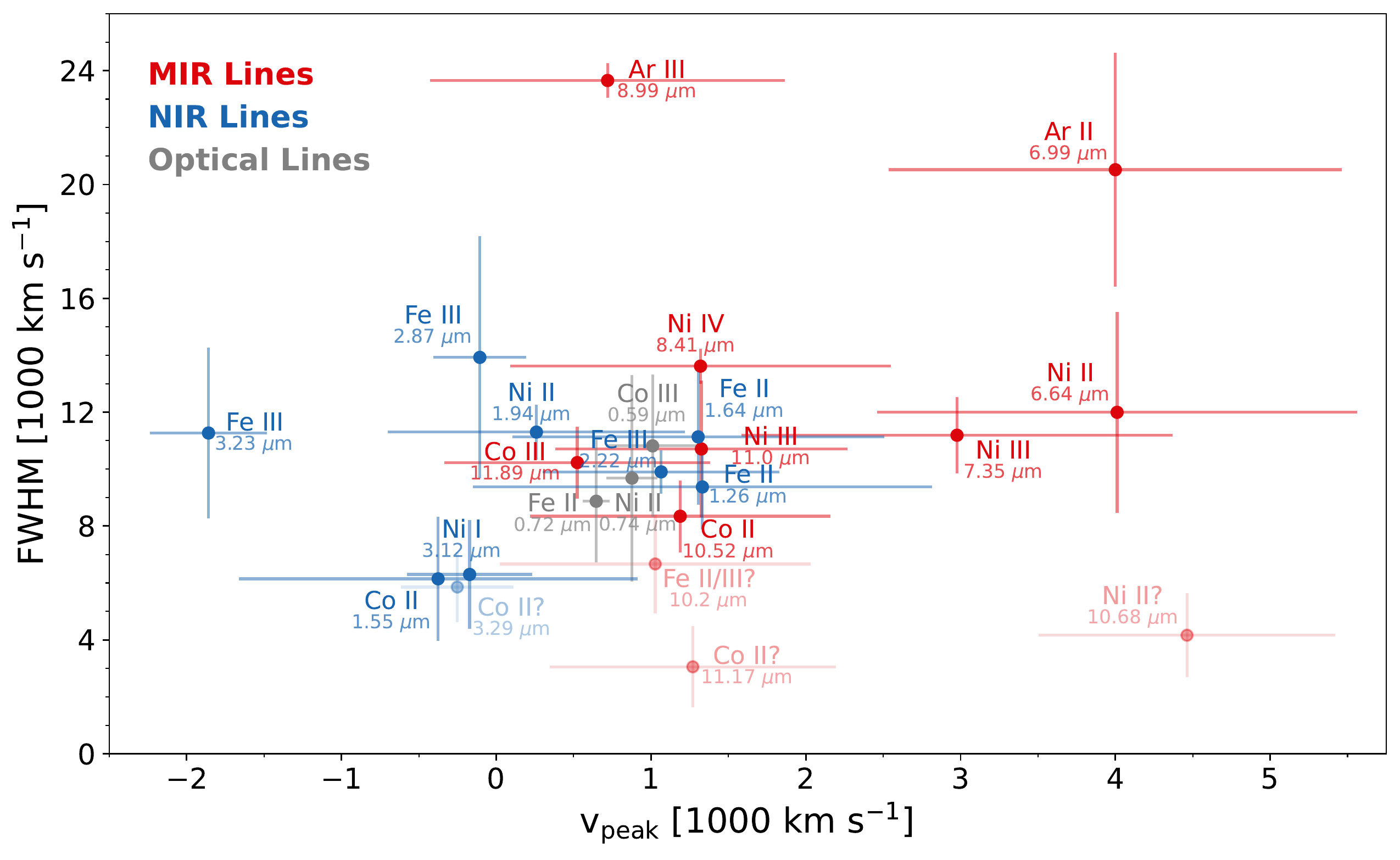}
    \caption{FWHM versus kinematic offset for species with fitted line profiles in \autoref{sec:profiles}. MIR lines (red) are compared against NIR lines (blue) and optical lines (grey). Error bars show fit uncertainties, in combination with wavelength calibration uncertainties for the MIR lines, and take into account instrument resolution. Low opacity points represent uncertain line identifications. Values and errors for these measurements are given in \autoref{tab:lines}.}
    \label{fig:FWHM_vels}
\end{figure*}

Appearing only in the MIR, the argon lines reveal new structure in the SN emission. \citet{Gerardy2007} find that the [Ar II] 6.985\um\ and [Ar III] 8.991\um\ lines have distinctive forked profiles in both SN~2005df and SN~2003hv, indicating an asymmetric, ring-shaped argon distribution. In contrast, for SN~2021aefx, the [Ar III] 8.991\um\ feature in SN~2021aefx has a very distinctive boxy shape, indicative of a hollow uniform sphere, or thick shell, of emission. 

Shown in \autoref{fig:ar_vels}, [Ar III] 8.991\um\ has a very broad FWHM $\approx$ 23,700\kms with an inner shell radius corresponding to a minimum velocity of $v_\text{min} =$ 8700 $\pm$ 200\kms\ and an outer radius corresponding to $v_\text{max} =$ 13,500 $\pm$ 300\kms. The flat-top of the [Ar III] 8.991\um\ line is only slightly sloped and is more symmetric than in the other SNe observed by \textit{Spitzer} and GTC/CanariCam. Indeed, the [Ar III] 8.991\um\ line shape exhibits considerable variation across the SNe in \autoref{fig:miri_ID}.

The [Ar II] 6.985\um\ feature in SN~2021aefx is heavily blended on both sides by [Ni II] 6.636\um\ and [Ni III] 7.349\um, making it difficult to conclusively determine its geometry. We find that the full $\sim$6.5--7.5\um\ feature is well fit by a sum of three Gaussian profiles, with the Gaussian representing [Ar II] 6.985\um\ having a very broad FWHM $\approx$ 20,600\kms\ and high redshifted $v_\text{peak} \approx$ 4000\kms. However, most lines from the same element share similar shapes, and we can fit the feature equally well with a boxy, shell profile for [Ar II] 6.985\um\ if we include an additional line which could be [Ni II] 6.920\um. In \autoref{fig:ar_vels} we show a fit to the [Ar II] 6.895\um\ line where we force it to have the same inner and outer shell radii as the [Ar III] 8.991\um\ fit. This forced fit reduces $v_\text{peak}$ of [Ar II] but yields a very high redshifted $v_\text{peak} \approx$ 5800\kms\ for [Ni II] 6.920\um. Without the boxy [Ar III] profile, we would not consider a boxy profile for [Ar II], and the fit is not unique since we could allow the [Ar II] shell to have different radial boundaries than [Ar III]. More detailed modeling of this feature is needed to fully disentangle the profile of [Ar II]. We further discuss the shapes of the Ar lines and their implications in \autoref{sec:conclusions}.

\section{Summary and Conclusions \label{sec:conclusions}}

We present nebular SN Ia NIR and MIR spectra of SN~2021aefx from \textit{JWST}, including the first observation of the 2.5$-$5\um\ region and the highest S/N MIR SN Ia spectrum to date. 

At the phase of our observations ($+$255d), all of the radioactive $^{56}$Ni has decayed away and the Ni emission lines that we see come from stable $^{58}$Ni. We unambiguously detect stable Ni in the ejecta of SN~2021aefx via the strong [Ni I] 3.120\um\ and [Ni II] 7.349\um\ lines, a clearly resolved [Ni IV] 8.405\um\ line, and several other weaker, blended Ni lines. Electron capture reactions producing stable iron-group elements require high density burning above $\sim$10$^8$ g cm$^{-3}$ that is found in carbon-oxygen white dwarfs with $M \gtrsim$ 1.2 $M_\odot$ \citep{Hoeflich1996, Iwamoto1999, Seitenzahl2013, Seitenzahl2017, Hoeflich2017}. The strong detection of stable Ni advocates for a massive, perhaps near-Chandrasekhar mass, progenitor for SN~2021aefx. However, our relatively low instrumental resolution does not allow us to probe the ejecta structure at the lowest velocities (the central region), and a more detailed quantitative analysis is needed to determine whether the stable nickel mass could be consistent with a lower-mass progenitor in a double detonation scenario \citep{Flors2020, Blondin2022}.

We fit emission-line profiles to prominent NIR and MIR lines to investigate their kinematic properties and geometric emission distributions. As shown in \autoref{fig:FWHM_vels}, the iron-group elements (Fe, Co, and Ni) largely cluster around a redshifted kinematic offset of $\sim$1000 $\pm$ 1000\kms\ and a FWHM of $\sim$ 11,000 $\pm$ 4000\kms. We find consistent kinematic offsets between the Co lines, with [Co III] having a slightly higher FWHM than [Co II], hinting that it may extend out to larger radii. This could indicate recombination in the higher density central region. \citet{McClelland2013} suggested that the more rapid decline seen in \textit{Spitzer} IRAC photometry of SN~2011fe at 3.6\um\ compared to 4.5\um\ could also be a result of recombination of doubly ionized species in the 3.6\um\ band; indeed, our data show [Fe III] emission should dominate that region.

The Ni lines have a slightly higher overall redshifted kinematic offset than Co and Fe. However, we caution that the largest redshifted kinematic offsets are found at the shortest MIRI wavelengths where the calibration is most uncertain. 

The reliably identified Fe lines are all consistent in FWHM. The [Fe III] 3.229\um\ line, which was not modeled with the SN~2015F model from \citet{Flors2020} like the other NIR Fe lines, is significantly blueshifted in kinematic offset compared to the other Fe lines; more detailed future modeling of this feature may bring it into closer agreement with the other Fe line measurements. Comparing the width of the [Fe II] 1.644\um\ line ($\sim$12,000\kms) to the models for SN~2014J from \citet[][e.g. see Figure 9]{Diamond2018}, we find that SN~2021aefx has a higher central density than SN~2014J. Though the uncertainties are large, there is a slight hint that a redshifted kinematic offset is seen for [Fe II] compared to [Fe III]; as described by \citet{Maeda2010}, the low-ionization lines trace the deflagration ash and this may be a signature of a delayed-detonation explosion where the initial deflagration produces an offset innermost ejecta while the subsequent detonation creates a spherically symmetric outer ejecta.

The argon lines are significantly broader than those of the iron-group elements, implying that argon extends to higher velocities, and correspondingly, radii. This result may support detonation models which produce stratified ejecta from nucleosynthesis in the low density outer layers leading to intermediate mass elements, and nucleosynthesis in the high density interior producing the iron group elements \citep{Khokhlov1991, Woosley1994, Wiggins1998, Fink2007, Sim2010}.

The flat-topped shape of the [Ar III] 8.991\um\ line indicates a thick shell geometry for the [Ar III] emission, which could either be produced by a physical lack of Ar in the central regions, or a lack of doubly ionized Ar in the interior. The [Ni IV] 8.405\um\ line, which has an ionization energy of 35.2 eV, has a smooth parabolic profile suggesting that [Ni IV] is present in the central region. Based just on ionization energies, it should then be possible to doubly ionize argon to [Ar III], which requires 27.6 eV, in the center. This would argue for a physical absence of Ar in the innermost ejecta, a flat-topped profile for [Ar II], and a stratified ejecta from a detonation. Comparison of ionization energies is likely too simplistic; indeed, \citet{Fransson2015} show that at late times in the high density regions, a large fraction of the energy comes from ionizations and excitations, rather than just thermal heating. A detailed future analysis should explore whether [Ni IV] can be formed in the central region without central emission of [Ar III].

Overall, the Gaussian, parabolic, and flat-topped shapes of the fit profiles for SN~2021aefx point to spherically symmetric distributions of emission for all species. This is consistent with the low level of continuum polarization of SNe Ia during the photospheric phase, as strong asymmetry of the radioactive distribution will lead to directional dependence of polarization \citep{Wang2008, Yang2022}. More \textit{JWST} MIR data of SNe Ia will improve our understanding of asymmetry in the explosions.

More detailed analyses, such as the derivation of elemental abundances and inferred masses, and modeling of this and future \textit{JWST} data of SN~2021aefx will be the subject of future work. Continuing observations of SN~2021aefx with \textit{JWST} will build a time series of four epochs of \textit{JWST} spectra and will be the most comprehensive SN Ia nebular IR dataset available. As part of \textit{JWST} Cycle 1 GO program 2072 (PI: S. W. Jha), another NIRSpec/Prism $+$ MIRI/LRS spectrum will be obtained roughly 100 days after the first, and two additional MIRI spectra of SN~2021aefx will be obtained through \textit{JWST} Cycle 1 GO program 2114 (PI: C. Ashall), including planned MIRI/MRS spectroscopy doubling the wavelength range out to 28\um, with higher spectral resolution. Analysis of SN~2021aefx over time will reveal the evolution of radioactivity, ionization, and density structure of the ejecta. 

SN~2021aefx is an excellent first reference SN Ia for \textit{JWST} observations and the initial analysis that we present here highlights the promise of \textit{JWST} to be transformative for the study of nebular phase SN Ia.

\facilities{\textit{JWST}, SALT}

\software{Astropy \citep{astropycollaboration_astropy_2018}, 
Matplotlib \citep{hunter_matplotlib:_2007}, 
NumPy \citep{oliphant_guide_2006}, PyRAF \citep{Pyraf}, PySALT \citep{PySALT}, dust extinction \citep{karl_gordon_2022_6397654}, jdaviz \citep{jdadf_developers_2022_7255461}, jwst \citep{Bushouse_JWST_Calibration_Pipeline_2022}, UltraNest \citep{Buchner2021, johannes_buchner_2022_7053560}
}

\vspace{6pt}
We thank Shelly Meyett for her consistently excellent work scheduling these JWST observations, Greg Sloan for assistance with the MIRI LRS wavelength solution correction, and Glenn Wahlgren for help with the NIRSpec observations.

This work is based on observations made with the NASA/ESA/CSA JWST. The data were obtained from the Mikulski Archive for Space Telescopes at the Space Telescope Science Institute, which is operated by the Association of Universities for Research in Astronomy, Inc., under NASA contract NAS 5-03127 for JWST. These observations are associated with JWST program \#02072. Support for this program at Rutgers University was provided by NASA through grant JWST-GO-02072.001. 

The SALT data presented here were obtained with Rutgers University program 2022-1-MLT-004 (PI: Jha). We are grateful to SALT Astronomer Rosalind Skelton for taking these observations.

L.A.K. acknowledges support by NASA FINESST fellowship 80NSSC22K1599.

T.S. is supported by the NKFIH/OTKA FK-134432 grant of the National Research, Development and Innovation (NRDI) Office of Hungary, the J\'anos Bolyai Research Scholarship of the Hungarian Academy of Sciences and by the New National Excellence Program (\'UNKP-22-5) of the Ministry for Innovation and Technology of Hungary from the source of NRDI Fund.

Time-domain research by the University of Arizona team and D.J.S.\ is supported by NSF grants AST-1821987, 1813466, 1908972, and 2108032, and by the Heising-Simons Foundation under grant \#2020-1864.

K.M. acknowledges support from the Japan Society for the Promotion of Science (JSPS) KAKENHI grant JP18H05223 and JP20H00174. 

L.G. acknowledges financial support from the Spanish Ministerio de Ciencia e Innovaci\'on (MCIN), the Agencia Estatal de Investigaci\'on (AEI) 10.13039/501100011033, and the European Social Fund (ESF) ``Investing in your future'' under the 2019 Ram\'on y Cajal program RYC2019-027683-I and the PID2020-115253GA-I00 HOSTFLOWS project, from Centro Superior de Investigaciones Cient\'ificas (CSIC) under the PIE project 20215AT016, and the program Unidad de Excelencia Mar\'ia de Maeztu CEX2020-001058-M.

K.M. and M.D. are funded by the EU H2020 ERC grant \#758638.

J.P.H. acknowledges support from the George A.\ and Margaret M.\ Downsbrough bequest.

The Aarhus supernova group is funded in part by an Experiment grant (\#28021) from the Villum FONDEN, and by a project 1 grant (\#8021-00170B) from the Independent Research Fund Denmark (IRFD).

C.L. acknowledges support from the National
Science Foundation Graduate Research Fellowship under Grant No. DGE-2233066

A.F. acknowledges support by the European Research Council (ERC) under the European Union’s Horizon 2020 research and innovation program (ERC Advanced Grant KILONOVA No. 885281).

\bibliography{zotero_abbrev}

\end{document}

%% file: affil.tex
\newcommand{\LCO}{\affiliation{Las Cumbres Observatory, 6740 Cortona Drive, Suite 102, Goleta, CA 93117-5575, USA}}
\newcommand{\UCSB}{\affiliation{Department of Physics, University of California, Santa Barbara, CA 93106-9530, USA}}

\newcommand{\OKC}{\affiliation{Oskar Klein Centre, Department of Physics, Stockholm University, Albanova University Center, SE-106 91, Stockholm, Sweden}}

\newcommand{\STScI}{\affiliation{Space Telescope Science Institute, 3700 San Martin Drive, Baltimore, MD 21218-2410, USA}}

\newcommand{\UA}{\affiliation{Steward Observatory, University of Arizona, 933 North Cherry Avenue, Tucson, AZ 85721-0065, USA}}

\newcommand{\Carnegie}{\affiliation{Observatories of the Carnegie Institute for Science, 813 Santa Barbara Street, Pasadena, CA 91101-1232, USA}}

\newcommand{\IfA}{\affiliation{Institute for Astronomy, University of Hawai`i, 2680 Woodlawn Drive, Honolulu, HI 96822-1839, USA}}
\newcommand{\UCSC}{\affiliation{Department of Astronomy and Astrophysics, University of California, Santa Cruz, CA 95064-1077, USA}}

\newcommand{\Princeton}{\affiliation{Department of Astrophysical Sciences, Princeton University, 4 Ivy Lane, Princeton, NJ 08540-7219, USA}}

\newcommand{\JHU}{\affiliation{Department of Physics and Astronomy, The Johns Hopkins University, 3400 North Charles Street, Baltimore, MD 21218, USA}}

\newcommand{\GeminiNorth}{\affiliation{Gemini Observatory/NSF's NOIRLab, 670 North A`ohoku Place, Hilo, HI 96720-2700, USA}}

\newcommand{\Rutgers}{\affiliation{Department of Physics and Astronomy, Rutgers, the State University of New Jersey,\\136 Frelinghuysen Road, Piscataway, NJ 08854-8019, USA}}
\newcommand{\FSU}{\affiliation{Department of Physics, Florida State University, 77 Chieftan Way, Tallahassee, FL 32306-4350, USA}}

\newcommand{\TAMU}{\affiliation{Department of Physics and Astronomy, Texas A\&M University, 4242 TAMU, College Station, TX 77843, USA}}
\newcommand{\Mitchell}{\affiliation{George P.\ and Cynthia Woods Mitchell Institute for Fundamental Physics \& Astronomy, College Station, TX 77843, USA}}

\newcommand{\ICE}{\affiliation{Institute of Space Sciences (ICE, CSIC), Campus UAB, Carrer de Can Magrans, s/n, E-08193 Barcelona, Spain}}
\newcommand{\IEEC}{\affiliation{Institut d'Estudis Espacials de Catalunya, Gran Capit\`a, 2-4, Edifici Nexus, Desp.\ 201, E-08034 Barcelona, Spain}}

\newcommand{\MSUPA}{\affiliation{Department of Physics and Astronomy, Michigan State University, East Lansing, MI 48824, USA}}
\newcommand{\MSUCMSE}{\affiliation{Department of Computational Mathematics, Science, and Engineering, Michigan State University, East Lansing, MI 48824, USA}}
\newcommand{\UPitt}{\affiliation{Department of Physics and Astronomy and Pittsburgh Particle Physics, Astrophysics and Cosmology Center (PITT PACC), University of Pittsburgh, 3941 O’Hara Street, Pittsburgh, PA 15260, USA}}
\newcommand{\ICCUB}{\affiliation{Institut de Ciències del Cosmos (ICCUB), Universitat de Barcelona (IEEC-UB), Martí Franqués 1, E08028 Barcelona, Spain}}
\newcommand{\USzeged}{\affiliation{Department of Experimental Physics, Institute of Physics, University of Szeged, D\'om t\'er 9, Szeged, 6720 Hungary}}
\newcommand{\ELKHSZTE}{\affiliation{ELKH-SZTE Stellar Astrophysics Research Group, H-6500 Baja, Szegedi \'ut, Kt. 766, Hungary}}
\newcommand{\Konkoly}{\affiliation{Konkoly Observatory, Research Centre for Astronomy and Earth Sciences, E\"otv\"os Lor\'and Research Network (ELKH), Konkoly-Thege Mikl\'os \'ut 15-17, 1121 Budapest, Hungary}}
\newcommand{\Prague}{\affiliation{Astronomical Institute, Academy of Sciences, Bo\v{c}n\`{i} II 1401, Prague, Czech Republic}}
\newcommand{\AMNH}{\affiliation{Department of Astrophysics, American Museum of Natural History, Central Park West and 79th Street, New York, NY 10024-5192, USA}}
\newcommand{\UDublin}{\affiliation{School of Physics, Trinity College Dublin, The University of Dublin, Dublin 2, Ireland}}
\newcommand{\VTech}{\affiliation{Department of Physics, Virginia Tech, Blacksburg, VA 24061, USA}}
\newcommand{\USheffield}{\affiliation{Department of Physics and Astronomy, University of Sheffield, Hicks Building, Hounsfield Road, Sheffield S3 7RH, U.K.}}
\newcommand{\Aarhus}{\affiliation{Department of Physics and Astronomy, Aarhus University, Ny Munkegade 120, DK-8000 Aarhus C, Denmark}}
\newcommand{\UOklahoma}{\affiliation{Homer L. Dodge Department of Physics and Astronomy, University of Oklahoma, 440 W. Brooks, Rm 100, Norman, OK 73019-2061, USA}}
\newcommand{\Hamburg}{\affiliation{Hamburger Sternwarte, Gojenbergsweg 112, D-21029 Hamburg, Germany}}
\newcommand{\UFlorida}{\affiliation{Department of Astronomy, University of Florida, Gainesville, FL 32611 USA}}
\newcommand{\ICG}{\affiliation{Institute of Cosmology \& Gravitation, University of Portsmouth, Dennis Sciama Building, Burnaby Road, Portsmouth PO1 3FX, UK}}
\newcommand{\DiRAC}{\affiliation{DiRAC Institute, Department of Astronomy, University of Washington, Box 351580, U.W., Seattle, WA 98195, USA}}
\newcommand{\Liverpool}{\affiliation{Astrophysics Research Institute, Liverpool John Moores University, Liverpool,  L3 5RF, UK}}
\newcommand{\MaxPlanck}{\affiliation{Max-Planck Institute for Astrophysics, Garching, Germany}}
\newcommand{\GSI}{\affiliation{GSI Helmholtzzentrum f\"ur Schwerionenforschung, Planckstra\ss{}e 1, 64291 Darmstadt, Germany}}
\newcommand{\KyotoU}{\affiliation{Department of Astronomy, Kyoto University, Kitashirakawa-Oiwake-cho, Sakyo-ku, Kyoto, 606-8502. Japan}}
\newcommand{\OSU}{\affiliation{Center for Cosmology and Astroparticle Physics, The Ohio State University, 191 West Woodruff Ave, Columbus, OH 43215, USA}}
\newcommand{\LasCampanas}{\affiliation{Carnegie Observatories, Las Campanas Observatory, Casilla 601, La Serena, Chile}}
\newcommand{\NotreDame}{\affiliation{Department of Physics and Astronomy, University of Notre Dame, Notre Dame, IN 46556, USA}}
\newcommand{\ESASTScI}{\affiliation{European Space Agency, Space Telescope Science Institute, 3700 San Martin Dr., Baltimore, MD 21218, USA}}